\documentclass[a4paper,twocolumn,11pt]{quantumarticle}
\pdfoutput=1
\usepackage[utf8]{inputenc}
\usepackage[english]{babel}
\usepackage[T1]{fontenc}
\usepackage{amsmath}
\usepackage{hyperref}
\usepackage{amsthm}
\usepackage{graphicx}
\graphicspath{{Figures/}}
\usepackage{subfigure}
\usepackage{caption}
\usepackage{geometry}
\usepackage{setspace}
\usepackage{tikz}
\usepackage{lipsum}
\usepackage{amsfonts}

\usepackage{enumitem}
\usepackage{comment}
\usepackage{booktabs}
\usepackage{amssymb}

\begin{document}

\title{Transforming qubits via quasi-geometric
	approaches}

\author{Nyirahafashimana Valentine}
\email{nyirahafashimanav@gmail.com}
\affiliation{Institute for  Mathematical Research (INSPEM), University Putra Malaysia, 43400 UPM Serdang, Selangor, Malaysia}
\orcid{https://orcid.org/0000-0002-9849-8163}
\author{Nurisya Mohd Shah}
\email{risya@upm.edu.my}
\homepage{
}
\orcid{https://orcid.org/0000-0000-0000-0000
}

\affiliation{Department of Physics, Faculty of Science, University Putra Malaysia, 43400 UPM Serdang, Selangor, Malaysia}
\author{Umair Abdul Halim}
\affiliation{Centre of Foundation Studies in Science of Universiti Putra Malaysia,  43400 UPM Serdang, Selangor, Malaysia.}
\orcid{https://orcid.org/0000-0002-9849-8163}
\author{Sharifah Kartini Said Husain}
\affiliation{Department of Mathematics and Statistics, Faculty of Science, University Putra Malaysia, 43400 UPM Serdang, Selangor, Malaysia}
\author{Ahmed Jellal}
\email{a.jellal@ucd.ac.ma}
\affiliation{Laboratory of Theoretical Physics, Faculty of Sciences, Choua\"ib Doukkali University, PO Box 20, 24000 El Jadida, Morocco}
\affiliation{Canadian Quantum  Research Center,
	204-3002 32 Ave Vernon,  BC V1T 2L7,  Canada}

\maketitle
\begin{abstract}

 We develop a theory based on quasi-geometric (QG) approach to transform a small number of qubits into a larger number of error-correcting qubits by considering four different cases. 
 More precisely, we use the 2-dimensional quasi-orthogonal complete complementary codes (2D-QOCCCSs) and quasi-cyclic asymmetric quantum error-correcting codes (AQECCs) via quasigroup and group theory properties. We integrate the Pauli $X$-gate to detect and correct errors, as well as the Hadamard $H$-gate to superpose the initial and final qubits in the quantum circuit diagram. We compare the numerical results to analyze the success, consistency, and performance of the corrected errors through bar graphs for 2D-QOCCCs and AQECCs according to their characteristics. The difficulty in generating additional sets of results and counts for AQECCs arises because mapping a smaller initial number of qubits to a larger final number is necessary to correct more errors. For AQECCs, the number of errors that can be corrected must be equal to or less than the initial number of qubits. High error correction performance is observed when mapping 1-qubit state to 29-qubits to correct 5 errors using 2D-QOCCCs. Similarly, transforming 1-qubit to 13-qubits using AQECCs also shows high performance, successfully correcting 2 errors. The results show that our theory has the advantage of providing a basis for refining and optimizing these codes in future quantum computing applications due to its high performance in error correction.
 \end{abstract}

\section{Introduction}\label{section 1}

Quantum computing is an exciting new frontier with the potential to solve complex problems faster than classical computing. However, because quantum information is based on the quantum state, it is very sensitive and prone to errors and decoherence. 
This makes quantum computing research very challenging. To protect quantum information from errors, a number of methods are being implemented and studied, namely quantum error correction  (QEC) \cite{PhilosophicalTransactions, PhysicalreviewA52}. In QEC, quantum error correction codes (QECCs) are designed to encode quantum states in qubits (two-state quantum systems) that can mitigate the effects of errors or decoherence in a limited number of individual qubits, ensuring that such problems have minimal or no effect on the stored data.

The study of error-detecting and error-correcting codes represents a seminal work in the field of information theory, notably advanced by Shor~\cite{PhysicalreviewA52}. This theory, which was first proposed over seven decades ago, has since had a profound impact on the field of classical computing. As the field of quantum information continues to grow and yield intriguing results, the theory of error detection and correction codes is also being revisited with the aim of successfully implementing a quantum version.
This work thus examines orthogonal geometric (OG) approaches in QEC with the objective of elucidating the more generalized properties of error-correcting codes (ECCs). Calderbank {\it et al.} seminal work~\cite{PhysRevLett} examines the profound impact of OG approaches on the advancement and efficacy of QEC. The framework provides a simplified representation of QECCs, enabling the generation of new code examples. These codes represent different configurations, including those with three to eight qubits correcting one error, four to ten qubits correcting one error, one to thirteen qubits correcting two errors, and one to twenty-nine qubits correcting five errors. This construction is based on the structural properties of certain subgroups, including $O(2^n)$ and $U(2^n)$~\cite{ProceedingsLondon}.
The work by Calderbank {\it et al.}~\cite{PhysRevLett} establishes a link between QECCs in Hilbert space and binary OG. However, the geometric aspect of this work lacks a comprehensive theoretical and mathematical formalism concerning bounds on the minimal distance of the constructed codes, which requires further theoretical analysis. Despite this, Calderbank's investigation represents a foundational contribution, offering valuable insights into the role of OG principles in driving continued progress in QEC.

Notable contributions focused on the construction of quasi-orthogonal complete complementary codes (QOCCCs) through the interplay of OG and group theory (GT). Dávideková's study extends the construction method for 2D-QOCCCs by extending insights from 1D-complete complementary codes (1D-CCCs)~\cite{ResearchPapers} . This generalization increases the versatility and applicability of QOCCCs in QEC. The two papers~\cite{IEEEWireless} and~\cite{IEEETransactionscommunications} have both made significant contributions to the development of quasi-orthogonal (QO) codes, which play a critical role in minimizing interference and errors in communication systems. The first paper~\cite{IEEEWireless} introduces an innovative rotation-based method to improve the performance of QO codes, while the second~\cite{IEEETransactionscommunications} focuses on the design of rate-one QO codes with partial diversity. Building on these efforts, the paper~\cite{IEEETransactionsInformation} contributes to the improvement of quasi-orthogonal space-time block codes (QOSTBCs) by proposing signal constellations that ensure both full diversity and fast maximum probability decoding. Taken together, these studies advance the understanding and application of QO codes, which are critical for ensuring reliable data transmission in the presence of noise and interference. Turcsány's innovative approach~\cite{ieeexplore.ieee.org} introduces a groundbreaking family of 2D-QOCCCs that provide an increased number of signatures for a given number of elements. This contributes to the diversification of QOCCC constructions and provides increased flexibility in QEC.

Lv's study focuses on the construction of asymmetric quantum error-correcting codes (AQECCs) from quasi-cyclic (QC) codes~\cite{7}. By presenting explicit constructions and computational results for binary and ternary AQECCs, this research addresses the asymmetry in QEC with practical solutions. Several research efforts have investigated the design of AQECCs using QC codes and group theory. The paper~\cite{7} contributes to this exploration by presenting a family of r-generator QC codes and their dual codes and subsequently using them in the construction of AQECCs. The paper~\cite{arXivpreprint} presents methods for generating asymmetric quantum cyclic codes, including BCH, RS, and RM codes. This involved the use of generator polynomials and the definition of sets of classical cyclic codes. However, the concept of asymmetric entanglement-assisted quantum error-correcting codes (asymmetric EAQECCs) research was introduced and extended in~\cite{IEEEAccess}.

In this work, we propose a theory to transform an $N$-qubit quantum representation into an $M$-qubit form capable of correcting $P$ errors using the quasi-orthogonal under the 2D-QOCCCs and quasi-cyclic over the AQECCs concept. This concept extends the idea of Caldebank's 1997 generalized construction of new codes using orthogonal geometry in QEC. We simulate to learn more about special codes called 2D-QOCCCs and AQECCs to understand them using two important ideas: quasi-orthogonality (QO) or quasi-cyclic (QC) and GT properties. The results of the experiment and the pick quantum states, which have $'0'$ and $'1'$ in their binary codes, follow the properties of GT and QO or QC. We present the comparison between the properties of QO or QC with GT. Furthermore, we show the performance of quantifying the effectiveness of error correction according to the code error rate.

The paper is structured as follows. Section~\ref{section 2} provides an overview of some GTs and their applications in the field of quantum computing. 
Section~\ref{section 3}  introduces the methodology, including the main details of the theoretical analysis and the mathematical formalism of 2D-QOCCCs and AQECCs. Section~\ref{section 4} is devoted to the numerical results throughout the Qiskit IBM quantum of four cases of transforming a small number of qubits into a larger number of qubits with the ability to correct multiple errors. Section~\ref{section 5} concludes our work.


\section{Group theory approach}\label{section 2}

In this section we present an overview of group theory, followed by a discussion of its application to quantum computing.

\subsection{Overview}

In linear algebra and group theory, special subgroups \(O(2^n)\) and \(U(2^n)\), also known as the Clifford group~\cite{ProceedingsLondon}, are essential for understanding the properties of orthogonal and unitary matrices over the field of complex or real numbers. The orthogonal group \(O(2^n)\) consists of matrices \(A\) of size \(2^n \times 2^n\), which satisfy the condition \(A^T A = I\), where \(A^T\) is the transpose and \(I\) is the identity matrix. For real matrices this implies \(A^T = A^{-1}\). Mathematically, it is written as 
\begin{equation}    
O(2^n) = \{ A \in \mathbb{R}^{2^n \times 2^n} \mid A^T A = I \}.
\end{equation}
The unitary group \(U(2^n)\) contains matrices \(U\) of the same size with the condition \(U^* U = I\), where \(U^*\) denotes the conjugate transpose. This can be expressed as
\begin{equation}
U(2^n) = \{ U \in \mathbb{C}^{2^n \times 2^n} \mid U^* U = I \}.
\end{equation}
Understanding these groups is crucial for exploring the role of orthogonal and unitary matrices in quantum computing, which is the subject of the present study.

Let \(P_n\) denote the conventional $n$-qubit Pauli group, defined as the subgroup of the Clifford group (\(O(2^n)\) and \(U(2^n)\)) generated by n-tensors of the single-qubit, bit-flip, and phase-flip Pauli operators \(X\), \(Z\), and \(iI\). Specifically for single qubits \cite{SpringerBerlinHeidelberg}, we have \(X|0\rangle = |1\rangle\), \(X|1\rangle = |0\rangle\), \(Z|0\rangle = |0\rangle\), and \(Z|1\rangle = -|1\rangle\). The corresponding n-qubit operators are expressed as \(X_1 = X \otimes (I \otimes (n-1))\), \(X_2 = I \otimes X \otimes (I \otimes (n-2))\), and so on, where \(I_m\) represents the identity operator on \(\mathbb{C}^m\), or simply \(I\) when the context is obvious.

In quantum computing and information processing, the special unitary group \(SU(2^n)\) and the special orthogonal group \(SO(2^n)\) are distinct from the term "Clifford group", despite their relevance in quantum computing, especially in the context of Clifford gates~\cite{25}. These groups are important because of their properties and operations within quantum systems. \(SU(2^n)\) consists of all unitary matrices \(U\) of size \(2^n\) with determinant 1, denoted as
\begin{equation}
SU(2^n) = \{ U \in \mathbb{C}^{2^n \times 2^n}\mid U^\dagger U = I, \text{det}(U) = 1 \}.\end{equation} On the other hand, the special orthogonal group \(SO(2^n)\) comprises all orthogonal matrices \(A\) of size \(2^n\) with determinant 1, defined as 
\begin{equation}
SO(2^n) = \{ A \in \mathbb{R}^{2^n \times 2^n} \mid A^T A = I, \text{det}(A) = 1 \}.\end{equation} 
These groups contribute significantly to the theoretical and practical aspects of quantum computing.
   The matrices in \(SO(2^n)\) symbolize rotations and reflections in Euclidean space, which have applications in geometry, physics, and computer science.
   The study ~\cite{26} contributes to the understanding of quantum operations and geometric transformations within the quantum computing landscape. 

In group theory a cyclic group has a special status, characterized by a single generator~\cite{cyberleninka.ru}. Within the structure of a group \(G\), the presence of an element \(a\) is fundamental in defining a cyclic group. If every element in \(G\) can be expressed as a power of \(a\), \(G\) transforms into a cyclic group denoted by \(G = \langle a \rangle\), with \(a\) serving as the generator.
We consider \( \langle g \rangle \) as a cyclic group generated by an element \( g \), defined as \( \langle g \rangle = \{ g^k \mid k \in \mathbb{Z} \} \), where \( g^k \) is the \( k^{th} \) power of \( g \). Furthermore, if \( g \) has a finite order \( n \), the cyclic group \( C_n \) can be expressed as 
\begin{equation}
 C_n = \{ g^0, g^1, g^2, \ldots, g^{n-1} \}.
 \end{equation} 
 This survey introduces the concept of cyclic groups and their importance in group theory.

   The operation in cyclic groups is conventionally represented multiplicatively. For a cyclic group \(G = \langle a \rangle\), the group operation is expressed as \(a^i \cdot a^j = a^{i+j}\). Notable properties of cyclic groups include a well-defined order, indicating the number of elements in the group, and the potential for being either finite, specifically in the cyclotomic quantum algebra of time perception~\cite{27}, or infinite, with the illustration that in an infinite array of \(2^n\) there are unitary solutions to the Yang-Baxter equation. The introduction of the work \cite{28} integrates quantum logic gates derived from cyclic groups \(C_n\) of arbitrary order $n$, using the quasi-triangular structure inherent in their group Hopf algebra.
   
A quasi-cyclic (QC) group has properties that combine those of both cyclic and infinite groups, allowing elements to be generated from a single element, similar to cyclic groups~\cite{AnnalidiMatematicaPuraedApplicata}. However, its structure does not strictly follow the pattern of an infinite cyclic group. The term "quasi" reflects this similarity to cyclic groups while allowing for variations in properties, such as cyclic patterns in group operations and the inclusion of shifts or translations within the group structure. 
Mathematically, a QC group \(G\) can be defined as a group for which there exists an element \(g\) such that each element of \(G\) can be expressed in the form \(g^n\)
\begin{equation}
\forall x \in G, \exists\, g \in G : x = g^n,
\end{equation}
where \(n\) is an integer.
In addition, the notation \( \text{QC}_n \) refers to a QC group of order \( n \), where its generator \( g \) need not be of infinite order, but still generates the entire group. Thus, its structure is similar to that of cyclic groups, with the possibility that the generator has finite order
\begin{equation}
\text{QC}_n = \{ g^0, g^1, g^2, \ldots, g^{n-1} \}.
\end{equation}

In other words, \(G\) may not be cyclic, but it has a cyclic-like property regarding the generation of its elements. In particular, it is essential to show the existence of a generator element \(g\) such that any element in G can be obtained by repeatedly applying the group operation to \(g\). For example, in illustrating a QC group of order 5, we represent all five generators in terms of a single element, denoted \(a\), and then confirm that \(a^5 = e\), where \(e\) is the identity element of the group.

\subsection{Quantum computing application}

In quantum computing, several mathematical groups play crucial roles, each contributing to different aspects of quantum operations and algorithms. The groups \(O(2^n)\) and \(U(2^n)\), subsets of the Clifford group, are fundamental in defining quantum gates. These include single-qubit gates, such as Pauli and Hadamard gates, and two-qubit gates, such as CNOT gates. In addition, these groups are central to the implementation of quantum error correction protocols, such as stabilizer codes, which ensure the reliability of quantum computation \cite{32}.


Note that the group \(SU(2^n)\) is essential for representing the evolution of quantum states, which is central to quantum algorithms and simulations. The group \(SO(2^n)\) is important for understanding geometric phases and the structure of quantum states, with applications in quantum information processing and quantum metrology \cite{33}.

QC groups are central to coding theory, error correction, and efficient communication system design. QC codes with cyclic-like properties detect and correct errors in telecommunications and data storage, ensuring reliable data transmission. They also improve network coding by supporting error correction in multi-path networks. They are being explored for QEC in quantum information processing, protecting quantum states from errors and noise, thus improving the stability of quantum computation \cite{Entropy25,IEEETransactionsonInformationTheory13}. Cyclic groups represent periodic phenomena in quantum systems and aid in group-based algorithms, while QC groups provide a framework for robust QECCs, exploiting their cyclic nature to achieve fault tolerance and robust error correction in quantum circuits \cite{34}.

\section{Methods}\label{section 3}
We introduce the idea of a subgroup and provide the mathematical formalism of 2D-QOCCCs and AQECCs for transforming $N$ qubits into $M$ qubits to correct $P$ errors, where $N$ is the number of initial qubits, $M$ is the number of final qubits, and $P$ is the number of errors corrected in the transformation. For the given QECCs, we focus on the following four cases, represented by $C_i$ ($i=1,\dots,4$)
\begin{enumerate}
\item[\textbf{$C_1$:}] Mapping 3-qubits to 8-qubits corrects for 1 error.
\item[\textbf{$C_2$:}] Mapping 4-qubits to 10-qubits corrects for 1 error.
\item[\textbf{$C_3$:}] Mapping 1-qubit to 13-qubits corrects for 2 errors.
\item[\textbf{$C_4$:}] Mapping 1-qubit to 29-qubits corrects for 5 errors.
\end{enumerate}

    \subsection{Introducing subgroups}
In quantum computing, the elements of \(SU(2^n)\) are essentially unitary matrices of size \(2^n\) with a determinant equal to 1. We construct this group by exploring quantum gates applicable to quantum computing scenarios. For example, the \(n\)-qubit Hadamard gate, denoted \(H\)-gate, which is a unitary matrix belonging to \(SU(2^n)\)
\begin{equation}
 H_{SU(2^n)} = \{ U \in SU(2^n) \mid U = H \otimes H \otimes \ldots \otimes H \},
 \end{equation}
with 
\(\otimes\) represents the tensor product. The subgroup \(H_{SU(2^n)}\) is formed by unitary matrices, each of which is a tensor product of $H$-gates acting on single qubits.
Note that the $H$-gate is mathematically represented as \cite{SpringerBerlinHeidelberg}
\begin{equation}
     H = \frac{1}{\sqrt{2}} \begin{bmatrix} 1 & 1 \\ 1 & -1 \end{bmatrix},
     \label{H}
     \end{equation}
which is a fundamental quantum gate in quantum computing and quantum information processing. When applied to a qubit, it transforms the basis states, creating a superposition. For example
\begin{align}
& H|0\rangle = \frac{1}{\sqrt{2}}(|0\rangle + |1\rangle),\\
& H|1\rangle = \frac{1}{\sqrt{2}}(|0\rangle - |1\rangle). 
\end{align}
Therefore, the $H$-gate can generate superposition, making it a crucial component in quantum circuits and various quantum algorithms.

    We construct a subgroup of \(SO(2^n)\), which is a subset of the special orthogonal group of dimension \(2^n\) that still satisfies all the properties of a group such as closure, associativity, and identity element, and inverse element.The elements of \(SO(2^n)\) are orthogonal matrices of size \(2^n\) with determinant 1. This subgroup is constructed by considering rotations in a particular subspace. Let \(R\) be a matrix representing a rotation in a \(k\)-dimensional subspace (\(1 \leq k \leq 2^n\)) of \(\mathbb{R}^{2^n}\). Then the collection of matrices generated by such rotations is a subgroup of \(SO(2^n)\) is
\begin{align}
H_{SO(2^n)} =& \{ R \in SO(2^n) \mid \\
 & \text{Rotation in }
 \, k\text{-dimensional subspace} \}.\nonumber 
\end{align}

A quasi-orthogonal (QO) subgroup $H_{SO(2^n)}$ appears as a distinctive subgroup within the framework of the special orthogonal group $SO(2^n)$. This subgroup is characterized by matrices belonging to $SO(2^n)$, which have a specific structure and exhibit defined, closely resembling orthogonal properties. Conceptually, these matrices symbolize rotations in a higher dimensional space, preserving certain geometric properties throughout the rotation process.
The motivation is to explore subgroups that exhibit quasi-orthogonal properties. One strategy, focusing on rotations near the identity matrix, leads to the construction of a subgroup called $H_{SO(2^n)}$ defined by
\begin{equation}\label{Subg}
H_{SO(2^n)} = \{ R \in SO(2^n) \mid R = I + \epsilon M \},
\end{equation}
where \(I\) is the identity matrix, \(\epsilon\) is a small positive parameter, and \(M\) is a skew-symmetric matrix that embodies a small rotation. The skew-symmetric property \(M^T = -M\) ensures purely imaginary eigenvalues. 
This subgroup \(H_{SO(2^n)}\) captures the concept of matrices closely related to identity since a small rotation corresponds to a subtle perturbation of the identity matrix in the context of \(SO(2^n)\). The intentional proximity of \(H_{SO(2^n)}\) to \(SO(2^n)\) is emphasized when \(\epsilon\) is small, providing a subgroup that retains QO properties.

In quantum computing, the construction of a QO subgroup \(H_{SO(2^n)}\) \eqref{Subg}
has implications for quantum gates, error correction, and system behavior under perturbations. We illustrate two examples of quantum gates related to the construction of 
\(H_{SO(2n)}\). The first is a single-qubit quantum gate ($H$-gate)~\cite{SpringerBerlinHeidelberg} where the QO subgroup construction \(R = I + \epsilon M\) can be applied to a single-qubit gate to create a superposition of qubits
   \begin{align}
  H = I + \frac{1}{\sqrt{2}}X, 
   \end{align}
  and \(X\) is the Pauli $X$-gate. The second is the two-qubit quantum gate (controlled $Z$-gate)~\cite{19} where the QO subgroup extends the construction to two qubits. The controlled $Z$-gate introduces entanglement between the qubits and is essential for several quantum computations 
   \begin{align}
   	 CZ =& I + \epsilon Z \otimes Z, \\
   	 = &\begin{bmatrix} 1 & 0 & 0 & 0 \\ 0 & 1 & 0 & 0 \\ 0 & 0 & 1 & \epsilon \\ 0 & 0 & \epsilon & 1 \end{bmatrix},
   \end{align}
   where \(Z\) is the Pauli $Z$-gate.

The QO subgroup \(H_{SO(2^n)}\) facilitates the study of quantum gates close to the identity, proving useful for certain quantum algorithms and ECCs. Its relevance in QEC is pronounced, helping in the study of noise effects on gates and in the design of robust ECCs. The consideration of the subgroup extends to numerical stability in quantum algorithms, especially in computations on quantum states and gates.~Matrices in \(H_{SO(2^n)}\), which act as small perturbations of the identity operator in Hilbert space \(\mathcal{H}\), provide insights into unitary operators that are crucial for preserving quantum state normalization.\newline

The QC group \( \text{QC}_n \) and the QO subgroup \(H_{SO(2^n)}\) belong to quasigroup theory~\cite{29}, which differs from group theory in some properties of the weak condition. QC and QO groups differ from classical group theory in fundamental aspects. A QO group can lack closure (\(ab \notin G\)), strict associativity (\((ab)c \neq a(bc)\)), and strict commutativity (\(ab \neq ba\)) for \( \exists \, a, b, c \in G \). Conversely, QC groups can exhibit irregularities in their cyclic behavior, potentially leading to deviations from predictable subgroup orders and generated subgroups. Formally, these deviations manifest themselves as the existence of elements \(a, b, c\) in a QO group \(G\) which violate closure, associativity, and commutativity.  The existence of elements \(g\) in a QC group \(G\), where the subgroup generated by \(g\), \(\langle g \rangle\), exhibits cyclic behavior with irregularities and may not conform to expected patterns of subgroup generation. 

 

When mapping a smaller number of initial qubits to a larger  number of final qubits, it becomes difficult to generate numerous result sets by using QC and GT techniques on AQECC to correct errors beyond the initial qubits.~The states are denoted as
\begin{equation}
 |\psi_{ij}\rangle = \begin{cases} 
\text{more outcomes} & \text{if}\quad i\geq j  \\
\text{two outcomes} & \text{if}\quad i<j  
\end{cases} 
\end{equation}
where $i$ is the number of initial qubits, while $j$ is the number of errors corrected in the transformation. Therefore, the number of errors corrected must not exceed the number of initial qubits to maximize the diversity of results. In QEC, increasing the error correction capability by increasing the number of physical qubits often comes at the cost of increased complexity and resource requirements. 

\subsection{QO code and GT transforming  qubits  for 2D-QOCCCs}\label{section 3.1}

In coding theory, the term "quasi-orthogonal (QO)" refers to specific properties of code constructions that are crucial for minimizing interference and errors when transmitting information over noisy channels. QO codes exhibit low cross-correlation between different sequences, a valuable property in communication systems such as Code Division Multiple Access (CDMA)~\cite{20,21}. In addition, these codes are
used in spread-spectrum communications, where signals are spread over a wide range of frequencies 
to increase resistance to interference.
This section aims to formulate a theory and mathematical framework dedicated to the transformation of an $N$-qubit quantum representation into an $M$-qubit state capable of correcting $P$ errors. This transformation is based on the use of the QO concept. Here, $N$ denotes the number of initial or logical qubits, $M$ denotes the final or physical qubit state, and $P$ denotes the number of correctable errors. Our proposed approach involves the use of the 2D-QOCCCs, which are derived from the 1D-CCCs with elements of length $N^2$~\cite{ResearchPapers}, to satisfy the requirements of this transformation.

\subsubsection{Organizing  1D-CCCs into a 2D array }

We begin by assembling a collection of 1D-CCCs, each characterized by a length of $N^2$, where $N$ is an integer. These codes are deliberately chosen for their advantageous cross-correlation properties, which are crucial for establishing quasi-orthogonality in the quantum coding framework. Let us symbolize this set of codes as $C$, with individual elements denoted by $c_i$, where $i = 0, 1, \ldots, N^2 - 1$. By using these codes, we lay the groundwork for the later development of 2D-QOCCCs.
Next, the elements $c_{ij}$ of $C$ are systematically organized into a 2D array of dimensions $N \times N$, where each element corresponds to an entry in the 1D-CCCs and is the coding representation for $N$-qubits. 
The quantum state for these $N$-qubits is initially represented as a linear combination
\begin{equation}
    |\psi\rangle = a|0\rangle + b|1\rangle + c|2\rangle +\dots+z|N\rangle,
\end{equation}
with the constant coefficients $a, b, c \dots z.$

For illustration, we represent a \(N \times N\) array with elements formed as $a_{ij}$
\begin{align}
a_{ij}=\begin{bmatrix}
   a_{11} & a_{12} & \ldots & a_{1N} \\
   a_{21} & a_{22} & \ldots & a_{2N} \\
   \vdots & \vdots & \ddots & \vdots \\
   a_{N1} & a_{N2} & \ldots & a_{NN}
\end{bmatrix}.
\end{align}
We improve QO by fine-tuning the array entries using the following general formula
\begin{equation}
    a_{ij}' = f(a_{ij}),
\end{equation}
where $a_{ij}'$ are the adjusted entries, $a_{ij}$ are the original entries, and $f(*)$ symbolizes a transformation function applied to each entry.  
To specifically address the goal of minimizing correlation and improving QO within the array, a method of reducing correlation between rows or columns is applied. This involves modifying entries $a_{ij}$ based on their influence on the overall correlation within the array, such as
\begin{align}
a_{ij}' = a_{ij} - \frac{\sum_{k=1}^{n} a_{ik} \cdot a_{jk}}{\sum_{k=1}^{n} a_{ik}^2}.
\end{align}

We proceed to expand the 2D array to improve the quasi-orthogonality. 
The tuning process for the 2D array is an integral step in refining the quasi-orthogonality of the quantum code. By precisely adjusting each \(c_{ij}\) value, we carefully manage a balance to reduce the connections between things without losing the effectiveness of the encoding method. However, modifying the array entries yields the result of a quantum state in terms of the adjusted quantum state  \(|\Psi\rangle\) for \(N\) qubits, which is 
\begin{equation}
    |\Psi\rangle = \sum_{i,j=1}^{N} 
    c_{ij}|i\rangle.
\end{equation}
This adjusted state encapsulates the refined encoding representation that integrates the QO adjustments made to the 2D array and contributes to the creation of a \(M\)-qubit quantum state \(|\Phi\rangle\)
\begin{equation}
    |\Phi\rangle = \sum_{i,j=1}^{N} 
    c_{ij}|i\rangle \otimes |j\rangle.
\end{equation}
The expansion of the quantum state reflects the increased dimensionality achieved by extending the encoding to \(M\)-qubits through tensor product operations between the initial qubits.

\subsubsection{Integrating error correction capability}

We integrate additional qubits to enable error detection and correction by creating a redundant $M$-qubit quantum state \(|\Phi_{\text{redundant}}\rangle\) and incorporating auxiliary qubits. In pursuit of a robust quantum information processing system, we introduce an error detection and correction mechanism within the 2D-QOCCCs. This crucial step increases the reliability of quantum encoding by introducing redundancy through the replication of selected qubits. We introduce a redundancy qubit to the \(M\)-qubit quantum state 
\begin{align}
    |\Phi_{\text{redundant}}\rangle &= |\Phi\rangle \otimes |k\rangle\\
    &= \sum_{i,j=1}^{N} 
     c_{ij}|i\rangle \otimes |j\rangle \otimes |k\rangle
    \label{qr1}
\end{align}
such that \(|k\rangle\) represents the redundant qubit introduced for error correction, \(|i\rangle\) and \(|j\rangle\) represent the basis states of the initial qubits. The summation extends over all values of \(i\) and \(j\), reflecting the incorporation of redundancy across the entire quantum state. This redundant state incorporates additional qubits (\(|k\rangle\)) to strengthen the quantum code against potential errors. 
%
The quantum state \(|\Phi_{\text{redundant}}\rangle\) of the system is expressed as a linear combination of basis states with coefficients derived from the updated elements of the 2D array. The coefficients \(c_{ij}\) are determined by the QO adjustments made during the expansion process. This representation is essential for understanding the probability amplitudes associated with each basis state, which has a profound effect on the collective state of the \(M\)-qubits system. It lays the foundation for subsequent analysis and manipulation within the framework of quantum information processing and quantum mechanics. The  probability amplitudes associated with each basis state of the coefficients \(c_{ij}\) is 
\begin{equation}
\text{Probability amplitude} = |c_{ij}|^2.
\end{equation}

\subsubsection{Error correction strategies and implementation}

There are two ways to handle error correction strategies, using either the quantum state representation or the matrix formalism.

\subsubsection*{3.2.3.1 Quantum state representation}
To implement error correction strategies, we use a combination of extra qubits, parity qubits, and auxiliary qubits. This approach is designed to build robust error correction techniques within the quantum system. The extra and parity qubits are introduced to establish error correction mechanisms. Subsequently, auxiliary qubits are incorporated to store redundant data, while additional parity qubits monitor the parity of qubit subsets. The quantum state representation incorporating redundancy, parity, and auxiliary qubits can be written as
\begin{align}
&|\Phi_{\text{redundant+parity+auxiliary}}\rangle =\\
&\sum_{i,j=1}^{N} 
d_{ij}|i\rangle \otimes |j\rangle 
\otimes |k\rangle \otimes |p\rangle \otimes |a\rangle.\nonumber
\end{align}
with specific coefficients \(d_{ij}\) and their arrangement depends on the properties and constraints of the 2D-QOCCC.
Then, using information from the auxiliary and parity qubits, we apply quantum error correction algorithms to identify and correct errors. 
The resulting quantum state representation of the corrected quantum state is 
    \begin{equation}
    |\Phi_{\text{corrected}}\rangle = \sum_{i,j=1}^{N} 
     d_{ij}'|i\rangle \otimes |j\rangle \otimes |k\rangle \otimes |p\rangle \otimes |a\rangle.
    \label{eq-1D-QOCCC}
\end{equation}
The application of quantum error correction algorithms, using information from auxiliary and parity qubits, can identify and correct errors. 

\subsubsection*{3.2.3.2 Matrix representation}
In our quantum encoding methodology, we explore the matrix representations that form the foundation of the entire process. For example,  let \(C\) denote the original 1D array of length \(N^2\), and \(A\) denote the corresponding 2D array of dimensions \(N \times N\), such as
\begin{align}
 &C = [c_0, c_1, \ldots, c_{N^2-1}]\\
&
   A = \begin{bmatrix} c_0 & c_1 & \ldots & c_{N-1} \\ c_N & c_{N+1} & \ldots & c_{2N-1} \\ \vdots & \vdots & \ddots & \vdots \\ c_{(N-1)N} & c_{(N-1)N+1} & \ldots & c_{N^2-1} \end{bmatrix}.
\end{align}
The adapted 2D array \(A'\) after quasi-orthogonal modifications is 
\begin{align}
 A' = \begin{bmatrix} d_{00} & d_{01} & \ldots & d_{0(N-1)} \\ d_{10} & d_{11} & \ldots & d_{1(N-1)} \\ \vdots & \vdots & \ddots & \vdots \\ d_{(N-1)0} & d_{(N-1)1} & \ldots & d_{(N-1)(N-1)}\end{bmatrix}.
\end{align}
These matrix representations succinctly capture the quantum states and their corresponding matrix transformations at each stage of the process. They provide a comprehensive visual understanding of the encoding process, including redundant, parity, auxiliary, and corrected states.

\subsection{QC code  and GT transforming qubits for AQECCs}\label{section 3.2}

Quantum computing holds immense potential to transform many computational tasks by processing information in a fundamentally different way than classical computers~\cite{22}. A notable advance in this area is the significant progress in ECCs, with a focus on asymmetric quantum error correction codes (AQECCs)~\cite{23}. AQECCs help to mitigate errors and increase the reliability of quantum systems. This work is essential for both the design and implementation of AQECCs and will contribute to the ongoing progress of quantum computing. The research on AQECCs is motivated by the overarching goal of transforming $N$-logical qubits into $M$-physical qubits while preserving the ability to correct $P$ errors. Integral to this transformative quest is a sophisticated mathematical framework based on the properties of QC codes and GT. This investigation aims to explain the intricacies involved in constructing codes that not only bridge the gap between logical and physical qubits, but also exhibit robust error correction functionalities.
\subsubsection{Dynamics of group action in quantum coding }
In the field of quantum coding, we denote \(\mathcal{Q}\) as a QC code generated by a set of shift operators \(g_i\), with \(i\) ranging from $0$ to \(M-1\). 
The representation of \(N\)-logical qubits  is encapsulated by the use of generators \(L_j\), with \(j = 1, 2, \ldots, N\). These generators serve as the fundamental elements that shape the landscape of logical information within the quantum coding paradigm. Within the intricate complexity of quantum coding, exploring the concept of group action involves considering a logical qubit \(L_j = (a_{1,j}, a_{2,j}, \ldots, a_{n,j})\), where \(n\) denotes the number of qubits embedded in the logical code. In this framework, we introduce a group action enabled by \(\{g_i\}\) on logical qubits
\begin{equation}
    g_i \cdot L_j = L_j^{(i)}. 
\end{equation}
This encapsulates the transformative essence of group action, where the application of \(g_i\) to \(L_j\) unfolds as an evolution denoted by \(L_j^{(i)}\). 
The operation \(a \mod b\) denotes the remainder of \(a\) divided by \(b\). In our context, \((i+1)\mod n\) manifests as a cyclic shift, which ensures that the transformation of the logical qubit obeys the circular nature of the operation
\begin{equation}
\begin{aligned}
&g_i \cdot (a_{1,j}, a_{2,j}, \ldots, a_{n,j}) = \\
(& a_{(i+1)\mod n, j}, a_{(i+2)\mod n, j},\ldots, a_{i, j}).
\end{aligned}
\label{17}
\end{equation}
This equation represents the encoding procedure and explains the use of the shift operation. The cyclic shifts bring a profound dynamism to the encoding process, illustrating how the logical qubit evolves under the influence of the group element \(g_i\).
\parskip 0.1in

\subsubsection{Error-correcting capability}
The error-correcting capability of the code, denoted by \(P\), is determined by its ability to detect and correct a given number of errors. This capability is inextricably linked to the minimum distance \(d\) of the code, which represents the smallest number of errors required to transform one valid codeword into another. Specifically, \(P\) is expressed in terms of the the floor function \cite{35}
\begin{equation}
    \begin{aligned}
        &P = \left\lfloor \frac{d-1}{2} \right\rfloor,\\
        & P \leq \frac{d-1}{2}< P+1,\\
        & 2P+1 \leq d<2P+3
        \end{aligned}
    \end{equation}
    The minimum distance \(d\) is the least error needed to change one valid codeword \([n,k,d]\) into another, then \(d\) must be an integer in the range \([2P+1,2P+2]\).
A codeword is a string of symbols, usually qubits in quantum coding, belonging to a particular set in a quantum error-correcting code. In this context, each codeword represents a valid encoded quantum state that is designed to be accurately retrieved even if errors occur during transmission or storage. The parameters \([n, k, d]\) describe the structure of the codewords, where \(n\) is the total number of physical qubits, \(k\) is the number of logical qubits, and \(d\) is the minimum distance. These parameters facilitate error detection and correction by allowing the receiver to identify and correct errors based on the minimum distance \(d\).

Here are illustrations for the four cases mentioned above:
\begin{itemize}
    \item Case 1 gives a minimum distance \(d = 3\), resulting in a \([8, 3, 3]\) code. The logical qubits and the physical qubits, respectively, are
    \begin{align*} 
     & |\psi \rangle_{3} = |000 \rangle +|001 \rangle +| 010 \rangle + \cdots +| 111 \rangle,  
     \\
   & |\phi \rangle_{8} =|00000000 \rangle +|00011111 \rangle \\
    &+| 00100101 \rangle + \cdots +|11111111 \rangle.
    \end{align*}
    \item Case 2 also gives \(d = 3\), forming a \([10, 4, 3]\) code. 
    The logical qubits and the physical qubits, respectively, are
      \begin{align*} 
  &|\psi \rangle_{4} =|0000 \rangle +|0001 \rangle + \cdots +|1111 \rangle, \\   
  &  |\phi \rangle_{10} =|0000000000 \rangle +| 0001111111 \rangle \\
    &+| 0010101010 \rangle + \cdots +| 1111111111 \rangle.
    \end{align*}
    \item Case 3 leads to \(d = 5\), creating a \([13, 1, 5]\) code. 
    The logical qubits and the physical qubits, respectively, are
      \begin{align*}
   &|\psi \rangle_{1} = | 0 \rangle +| 1 \rangle, \\
   & |\phi \rangle_{13} = | 0000000000000 \rangle + | 1111100000000 \rangle \\
    &+| 0000000000011 \rangle + \cdots + | 1111111111111 \rangle.
    \end{align*}
    \item Case 4 results in \(d = 11\), yielding a \([29, 1, 11]\) code. 
    The logical qubits and the physical qubits, respectively, are
     \begin{align*}
& |\psi \rangle_{1} = | 0 \rangle +| 1 \rangle, \\ 
    &| \phi \rangle_{29} =|00000000000000000000000000000 \rangle \\
   & +| 11111111111000000000000000000 \rangle \\
    &+\cdots +| 11111111111111111111111111111 \rangle.
    \end{align*}
\end{itemize}

For example, a \([8, 3, 3]\) code maps 3 logical qubits into 8 physical qubits with a minimum distance of 3, providing error correction. Similarly, a \([29, 1, 11]\) code maps 1 logical qubit into 29 physical qubits with a minimum distance of 11, providing robust error correction. These codes use algorithms to generate codewords that identify and correct errors based on their minimum distance.

In quantum error correction, encoding involves applying single-qubit gates (such as Pauli gates \(X, Y, Z\), Hadamard gate) to each logical qubit, two-qubit gates to create entanglement between pairs of physical qubits, and three-qubit gates for more complex mappings. Single-qubit gates modify individual qubit states to encode logical information. In contrast, two-qubit gates (such as CNOT gates) establish correlations between multiple physical qubits, creating entangled states that distribute this information. Three-qubit gates (such as the Toffoli gate) further enhance these mappings, facilitating the implementation of AQECCs \cite{36}. The goal is to distribute logical qubit data over more physical qubits while maintaining the ability to detect and correct errors, which is crucially determined by the minimum distance \(d\) of the code.

\subsubsection{Quantum error-correcting codes via group theory}
In the construction of QECCs, the code's strength is strengthened by applying GT concepts that create distinct subgroups. The formulation of error operators in the subgroup \(\mathcal{E}\), generated by the group \(\{g_i\}\), is complemented by the simultaneous consideration of logical operators in another subgroup \(\mathcal{L}\). 
The important stabilizer group \(\mathcal{S}\) materializes as the intersection of \(\mathcal{E}\) and \(\mathcal{L}\)
\begin{equation}
\mathcal{S} = \mathcal{E} \cap \mathcal{L} 
\end{equation}
which is a fundamental component in designing QECCs. Note that  $\mathcal{S}$ plays a critical role in mitigating errors and improving the reliability of quantum systems.

In achieving strong QEC, the imperative task is to identify check operators within the code \(\mathcal{Q}\). These operators have a peculiar property: they exhibit commutativity with logical operators while simultaneously exhibiting anticommutativity with error operators. Let \(L\) denote a logical operator, and \(E\) denote an error operator. This property is expressed by 
\begin{align}
\label{eq19a} & L \cdot S = S \cdot L \\
 & E \cdot S = -S \cdot E. 
\label{eq19b}
\end{align}
In equation \ref{eq19a} it is emphasized that stabilizer operators \(S\) and logical operators \(L\) work together smoothly, they commute. Conversely, in equation \ref{eq19b} it is shown that stabilizer operators \(S\) and error operators \(E\) tend to behave in opposite ways, they anti-commute.
This unique duality allows check operators to play their role in error detection without compromising the integrity of logical information. If we call \(C\) a check operator, the criteria for such an operator are clarified by 
    \begin{align}
   &     C \cdot L = L \cdot C \\
    &    C \cdot E = -E \cdot C.
    \end{align}
The specific characteristics and form of the check operators depend on the selected QC code and the intricate interplay between logical and error operators in the construction process.

\section{Numerical results and discussion}\label{section 4}

We present and discuss the numerical simulation and comparative analysis by the Qiskit IBM quantum of the transformation process from a small number of qubits to a larger number of qubits with the capability of error correction via the 2D-QOCCCs and AQECCs properties and group theory properties. Equation \ref{eq-1D-QOCCC} transforms $N$ logical qubit to produce $|\Phi\rangle$ and equation \ref{17} transforms $n\equiv N$ logical qubit. Then, we evaluate the performance, success, and reliability of the error-correcting process by statistical measurement. We consider the above four cases as mentioned in section \ref{section 3}.  


\subsection{2D-QOCCCs case}\label{4.1}

Here we present numerical and graphical simulations of 2D-QOCCCs of the quantum state representation referred to the equation \ref{eq-1D-QOCCC} and analyze them through the lens of two key concepts: group theory (GT) and quasi-orthogonality (QO) properties via the equation \ref{eq-1D-QOCCC}. The experimental results are consistent with the principles of GT and QO, especially for the presence of quantum states represented by $'0'$ and $'1'$ in their binary codes, respectively. A mathematical formalism is then used to combine these quantum states and to thoroughly investigate their frequency of occurrence and associated behavior. Bar charts are then generated to visualize the results. Due to the large volume of results and counts from the simulation data, we have chosen to randomly select samples for inclusion in our comparisons and analyses.



\begin{enumerate}
\item[\textbf{$C_1$:}]For mapping 3-qubits to 8-qubits, we can correct 1 error.

\begin{table}[!ht]
    \centering
    \caption{Simulation results of 2D-QOCCCs for both QO and GT to map 3-qubits to 8-qubits with one error correction. }
    \label{tab:1}
    \begin{tabular}{c|c}
         Outcomes (qubits)& Counts \\ \hline
        00010001 & 1\\
        00001001&3\\
        11110001&7\\
        11100100&2\\
        00000111&5\\
        00001001&3\\
        10010001&2\\
        11100100&2\\
        01000000&4\\
        10110101&3\\ \hline
    \end{tabular}
    \end{table}

\begin{figure}[!ht]
    \subfigure[3-qubits]{\includegraphics[scale=0.25]
    	{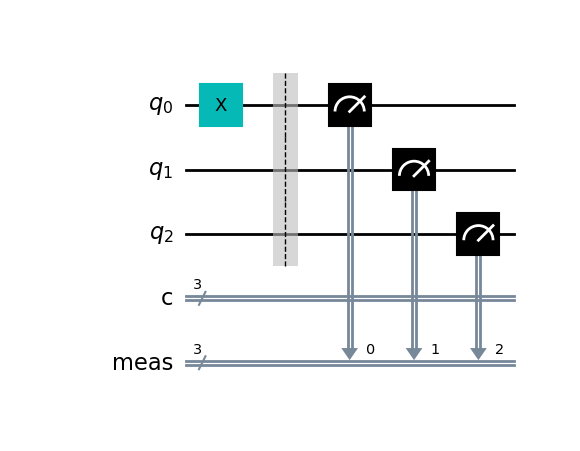}}
    \hspace{0.5cm} 
    \subfigure[8-qubits]{\includegraphics[scale=0.25]
    	{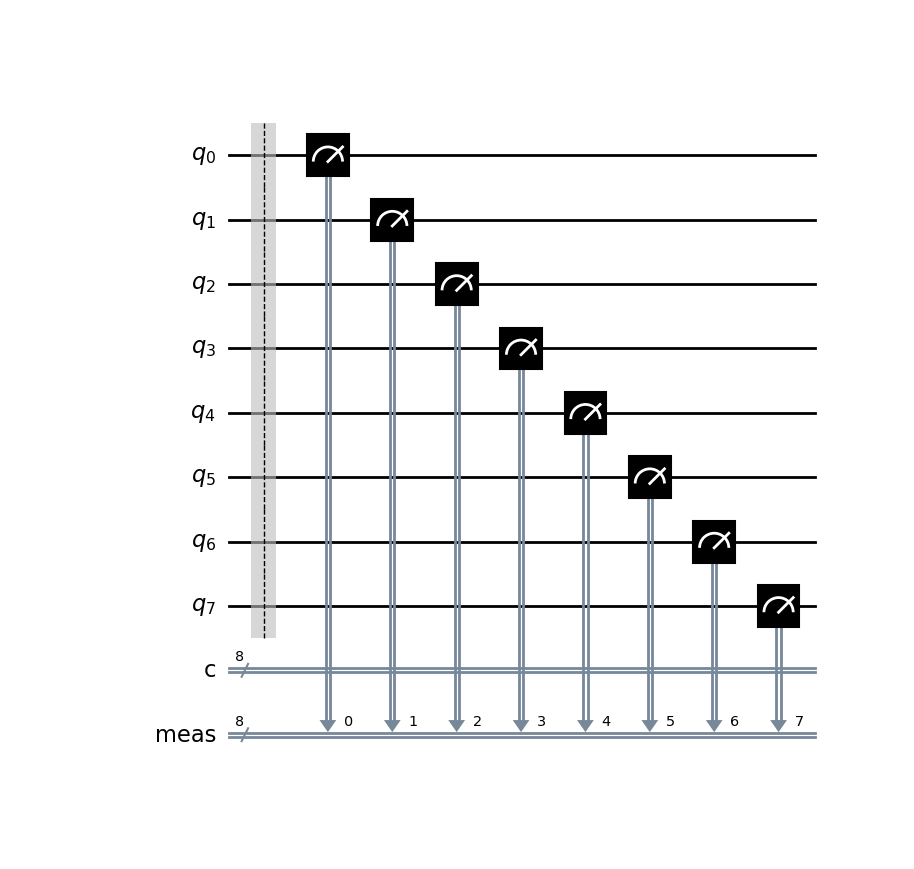}}
    \caption{Initial and final quantum states corresponding to 3-qubits and 8-qubits, with identified errors.}
    \label{fig:1}
\end{figure}

\begin{figure}[!ht]
    \centering
     \subfigure[QO (2D-QOCCCs) for 3-qubits to 8-qubits]{ \includegraphics[scale=0.30]{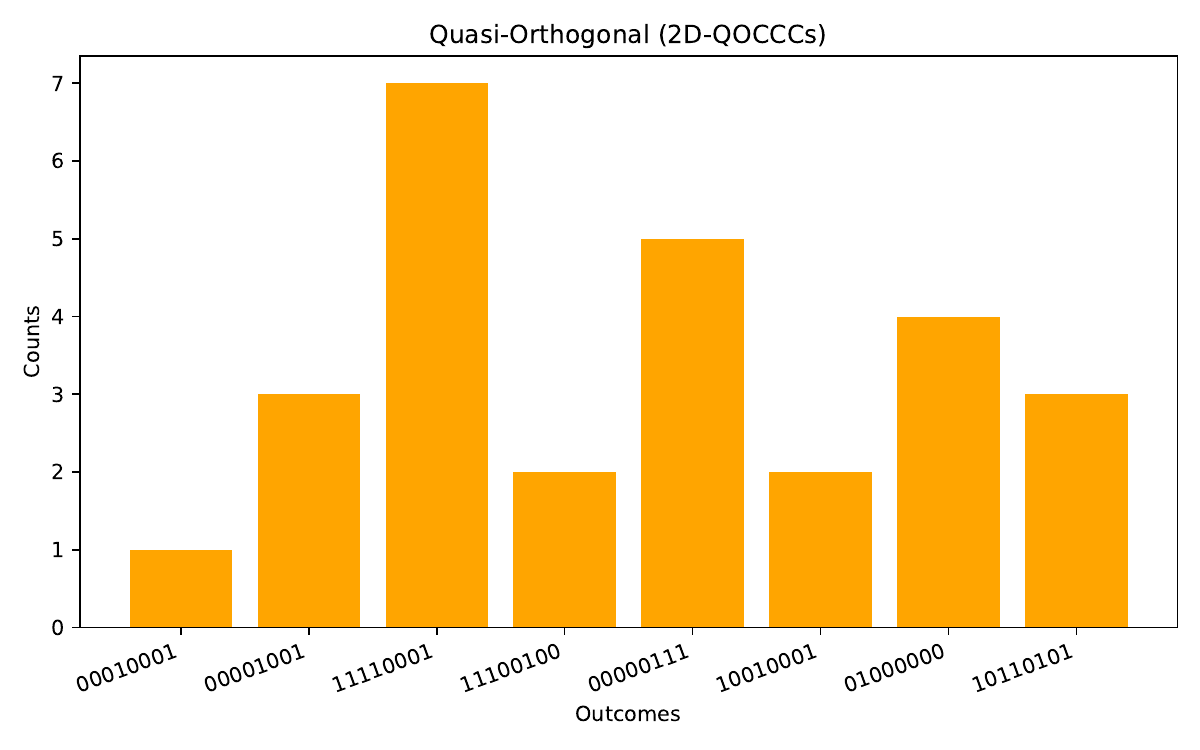}}
     \subfigure[GT (2D-QOCCCs) for 3-qubits to 8-qubits] {\includegraphics[scale=0.30]{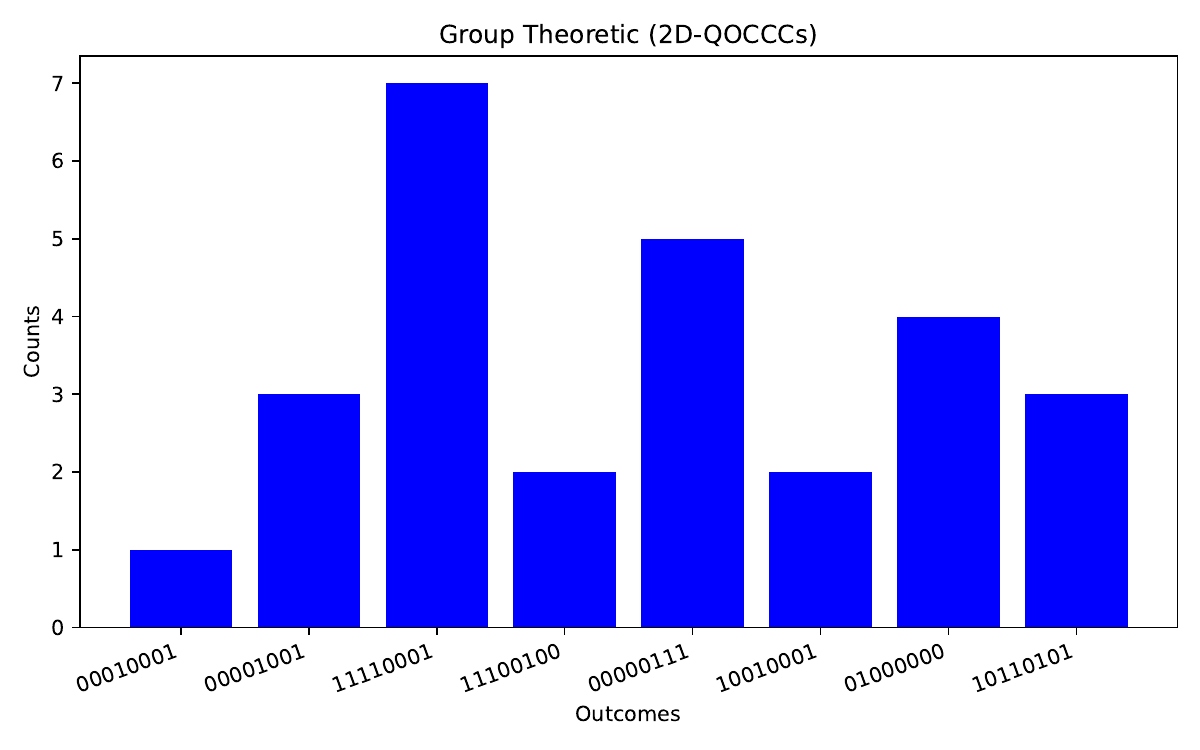}}
    \caption{Plots of GT and QO for 2D-QOCCCs to map 3-qubits to 8-qubits, with ability to correct an error.}
    \label{fig:2}
\end{figure}



Table~\ref{tab:1} shows the simulation results of 2D-QOCCCs for both QO and GT, showing the results and the number of counts. It is worth noting that the $11110001$ result has the highest count. 
This common result  indicates that one error has been corrected.

In Figure~\ref{fig:1}(a), which shows the initial 3-qubit diagram, the use of the $X$-gate to identify the error confirms that an error has been identified for correction, while Figure~\ref{fig:1}(b) shows the final 8-qubit diagram after incorporating parity and auxiliary qubits to correct the identified error.

Figure~\ref{fig:2} shows the plots of QO and GT for 2D-QOCCCs concerning the transformation of 3-qubits  into 8-qubits with the possibility of error correction. Figure~\ref{fig:2}(a) shows the plot for the results of 2D-QOCCCs, which are characterized by the presence of the binary digit $'1'$ in their representation. Figure~\ref{fig:2}(b) shows the results with $'0'$ in the binary code. As noted in Table~\ref{tab:1}, the presence of only one high-frequency result in these figures further attests to the correction of an error.

\item [\textbf{$C_2$:}] When mapping 4-qubits to 10-qubits, we still correct 1 error.

\begin{table}[!ht]
 \caption{Simulation results of 2D-QOCCCs for both QO and GT to map  4-qubit to 10-qubits, with one error correction.}
    \label{tab:2}
    \begin{tabular}{c|c}
        Outcome (qubits) & Counts \\ \hline
       0001000101  &1\\
       0000100101&3\\
       1111000101&7\\
       1110010001&2\\
       0000011101&5\\
       0000100101&3\\
       1001000101&2\\
       1110010001&2\\
       0100000010&4\\
       1011010101&3\\ \hline
       
    \end{tabular} 
\end{table}
\begin{figure}[!ht]
   \subfigure[4-qubits]{\includegraphics[scale=0.20]{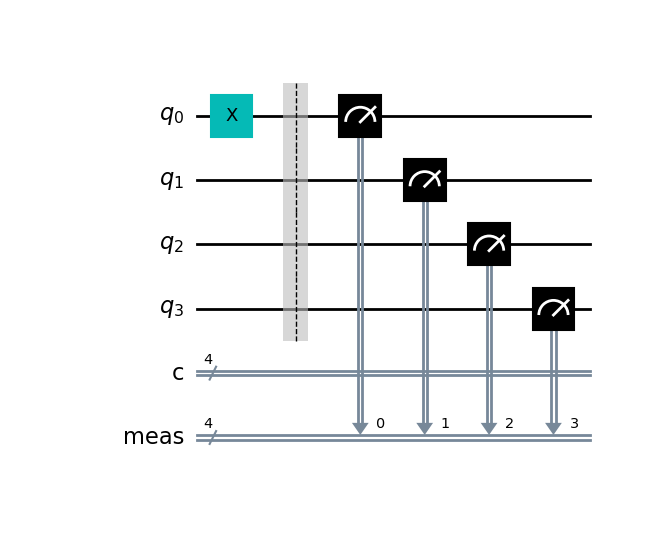}}
   \subfigure[10-qubits]{\includegraphics[scale=0.20]{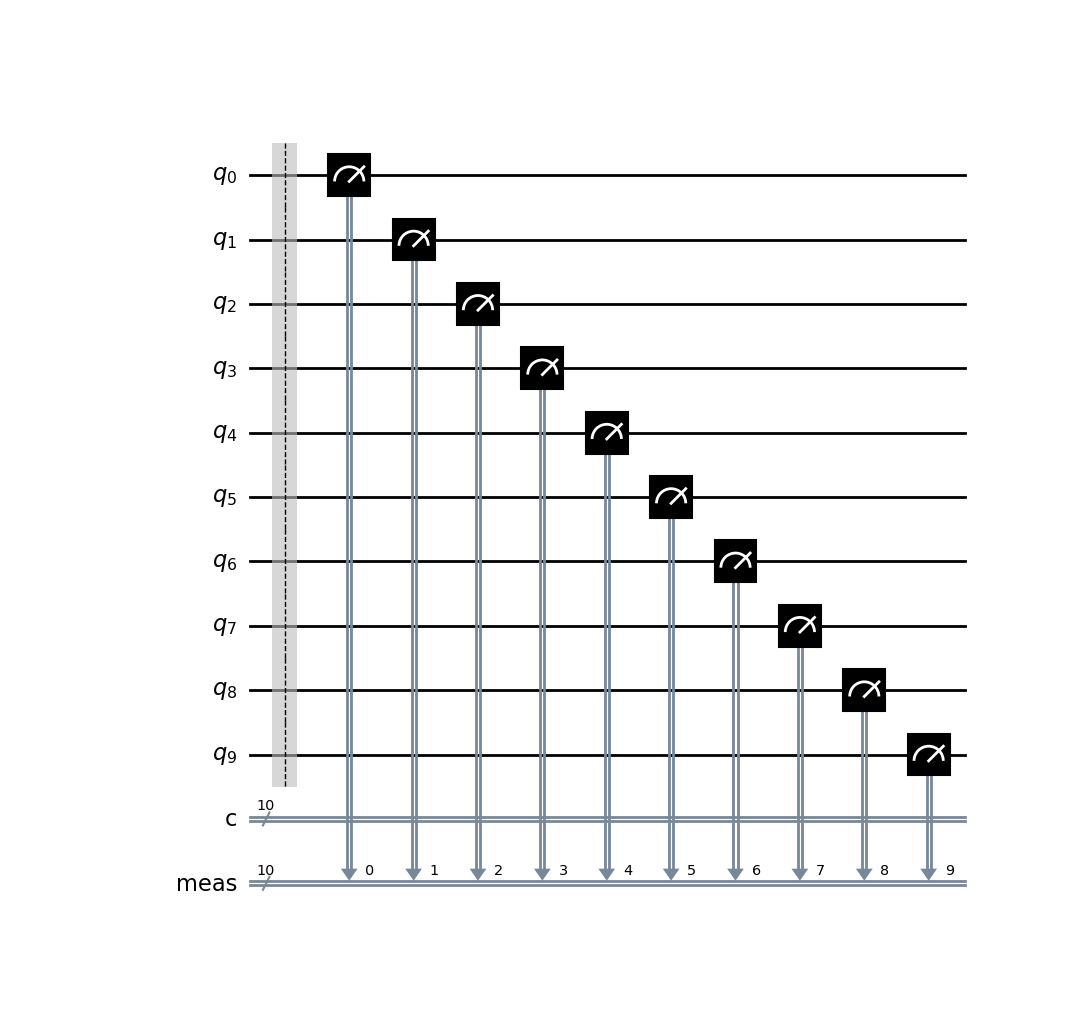}}
    \caption{
Initial and final quantum states corresponding to 4-qubits and 10-qubits, with identified errors.    
    }
    \label{fig:3}
\end{figure}

\begin{figure}[!ht]
     \subfigure[QO (2D-QOCCCs) for 4-qubits to 10-qubits]{ \includegraphics[scale=0.28]{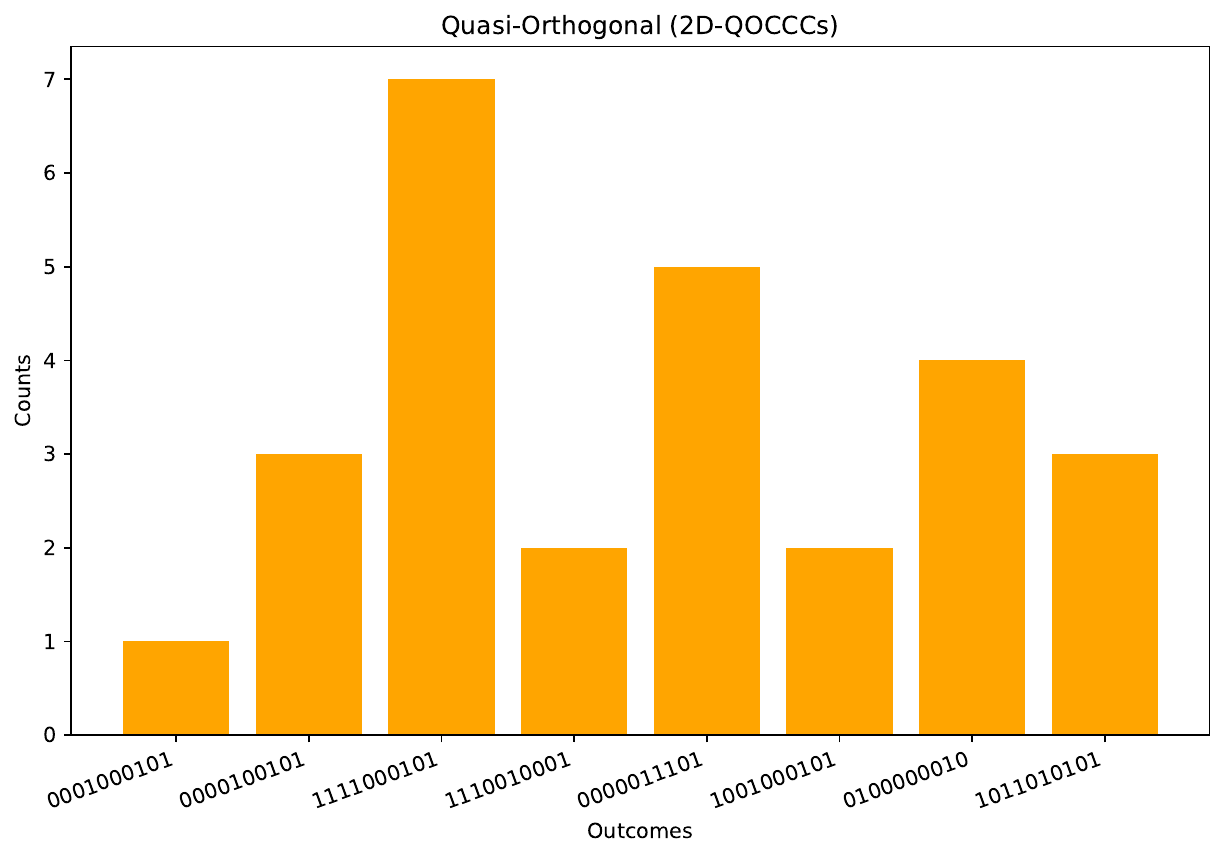}}
     \subfigure[GT (2D-QOCCCs) for 4-qubits to 10-qubits] {\includegraphics[scale=0.28]{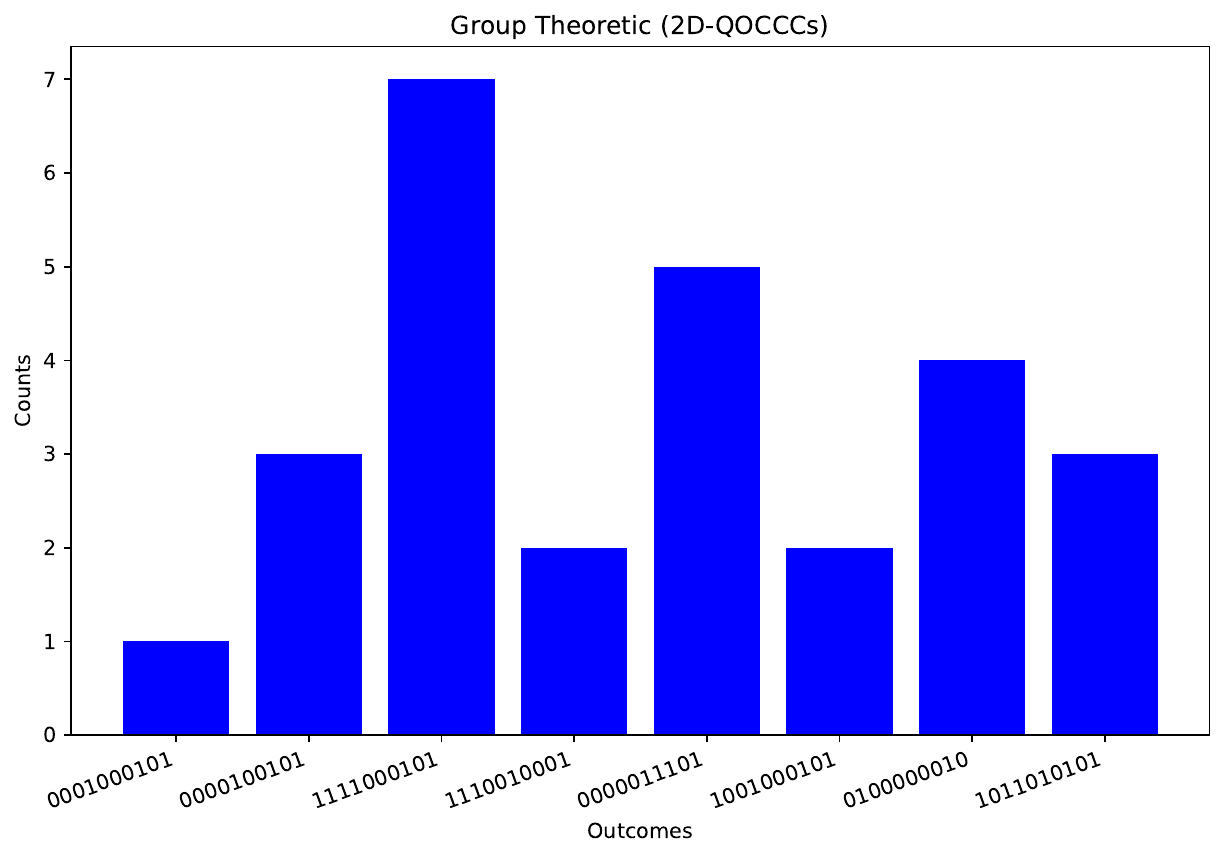}}
    \caption{Plots of GT and QO for 2D-QOCCCs to map 4-qubits to 10-qubits, with ability to correct one error.}
    \label{fig:4}
\end{figure}


Table~\ref{tab:2} shows the simulation results for 2D-QOCCCs, including both QO and GT along with their respective count frequencies. Notably, the outcome $1111000101$ stands out with the highest frequency, indicating its frequent recurrence. In contrast, other outcomes have lower counts, indicating lower accuracy. The prevalence of the high-frequency result suggests the successful correction of a single error.

Figure~\ref{fig:3}(a) shows the initial 4-qubit diagram, where the application of the $X$-gate confirms the identification of an error for correction. Subsequently, Figure~\ref{fig:3}(b) shows the final 10-qubit diagram after the integration of parity and auxiliary qubits to correct the identified error.

Figure~\ref{fig:4} shows the plots for QO and GT for 2D-QOCCCs to map 4-qubits to 10-qubits with error correction capabilities. Figure~\ref{fig:4}(a) shows results characterized by the presence of the binary digit $'1'$, while Figure~\ref{fig:4}(b) shows results with $'0'$ in the binary code. Notably, as indicated in Table~\ref{tab:2}, the prevalence of a single high-frequency result in both figures confirms the correction of an error. The similarity between the plots in Figure~\ref{fig:4}(a) and Figure~\ref{fig:4}(b) suggests comparable performance in the context of 2D-QOCCCs.


  \item [\textbf{$C_3$:}]When mapping 1-qubit to 13-qubits, we can correct 2 errors. 
\begin{table}[!ht]
    \centering
    \caption{Simulation results of 2D-QOCCCs for both QO and GT to map 1-qubits to 13-qubits with error correction.}
    \label{tab:1-13 qubits}
    \begin{tabular}{c|c}
       Outcome (qubits)  & Counts  \\ \hline
         0001000100010 & 1\\
         0000100100001& 3\\
         1111000100011& 7\\
          1110010001100& 7\\
          0000011100000&5\\
          0000100100001&3\\
          1001000100010&2\\
          1110010001100&1\\
          0100000010001&4\\
          1011010111011&3\\ \hline
    \end{tabular}
\end{table}
\vspace{3pt}
\begin{figure}[!ht]
    \centering
     \subfigure[1-qubit]{\includegraphics[scale=0.15]
     	{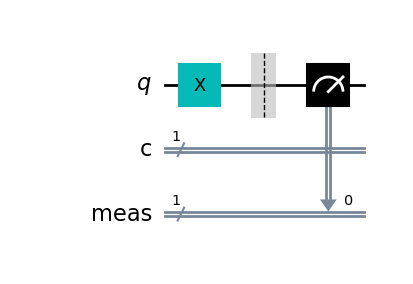}}
      \subfigure[13-qubits]{\includegraphics[scale=0.15]
   {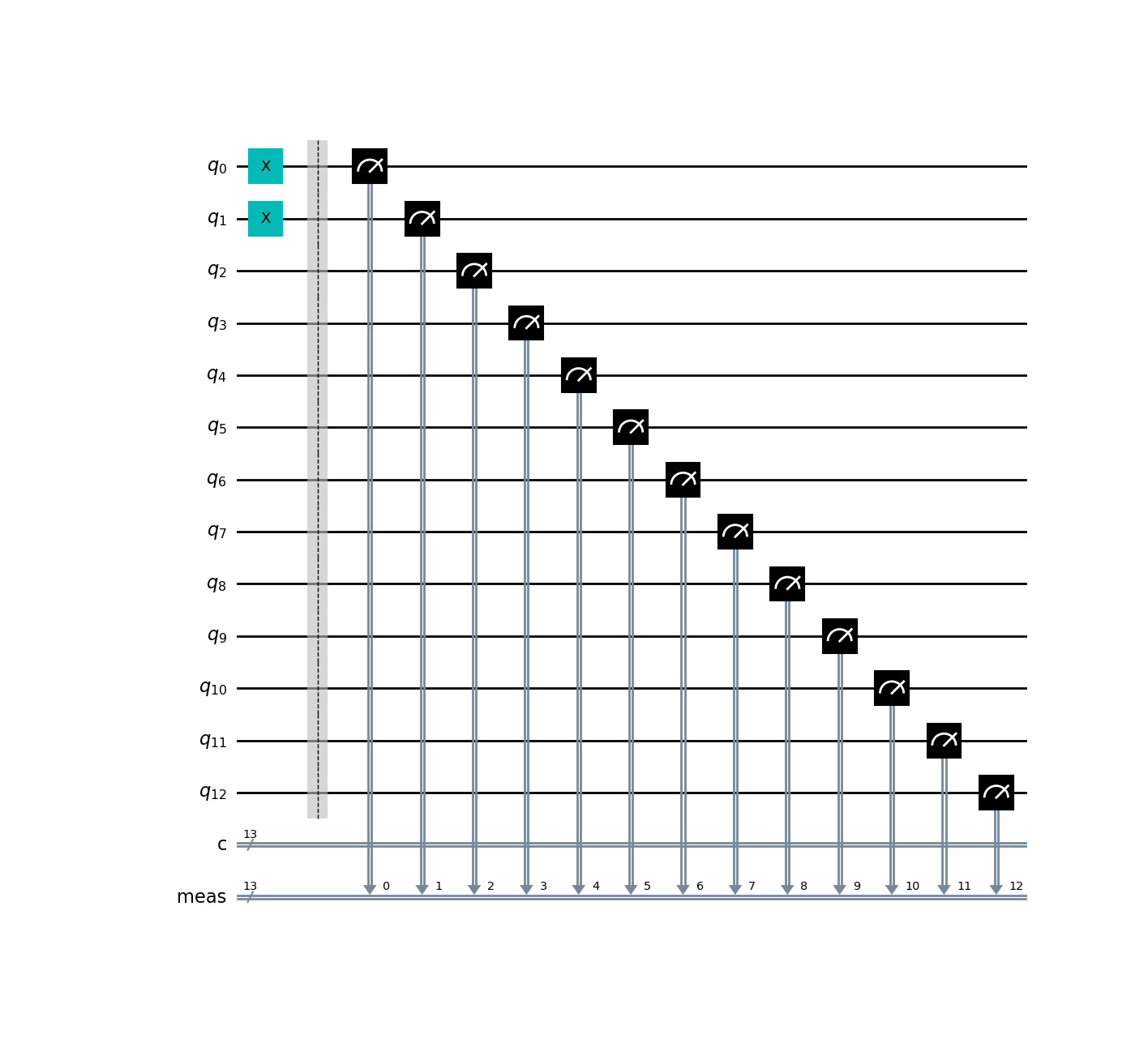}}
    \caption{Initial and final quantum states corresponding to 1-qubit and 13-qubits, with identified errors.}.
    \label{fig:5}
\end{figure}

\begin{figure}[!ht]
     \subfigure[\small{QO (2D-QOCCCs) for 1-qubit to 13-qubits }]{ \includegraphics[scale=0.25]{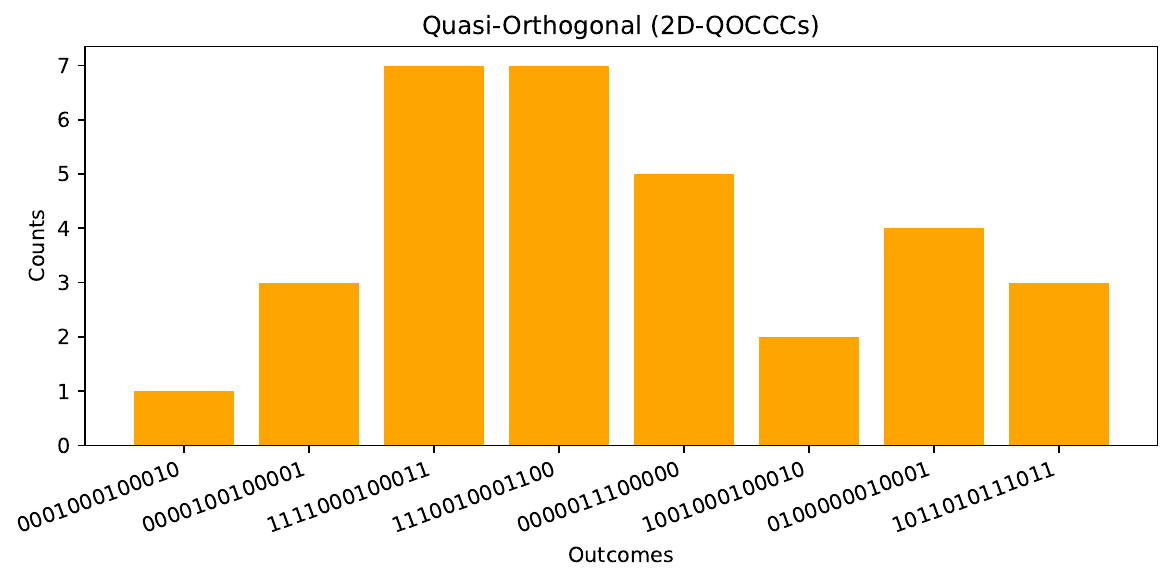}}
     \subfigure[\small{ GT (2D-QOCCCs) for 1-qubit to 13-qubits}] {\includegraphics[scale=0.25]{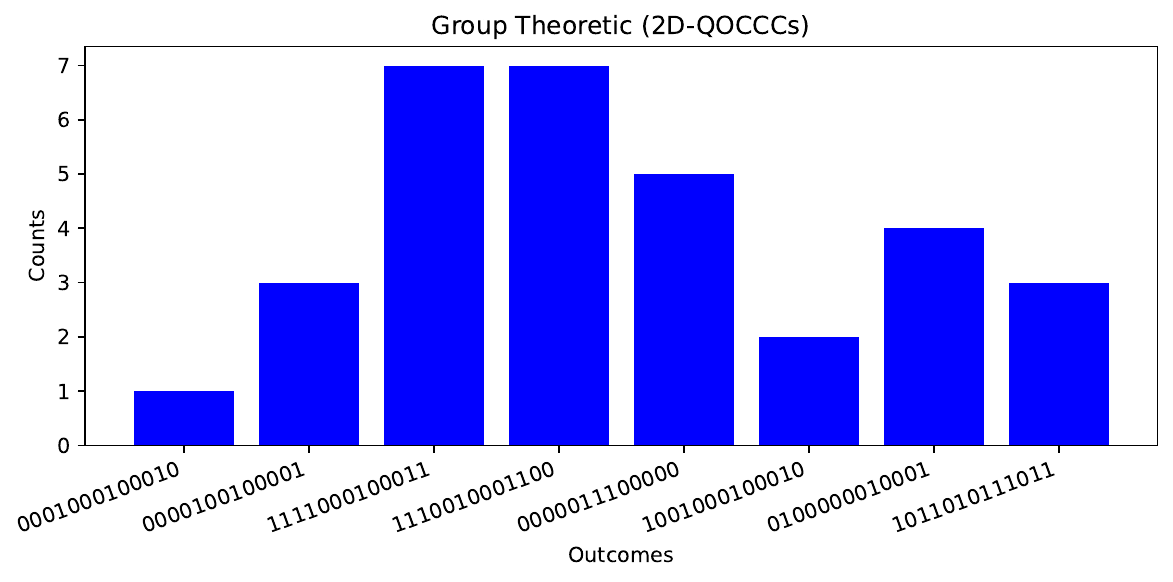}}
    \caption{\small{
Plots of GT and QO for 2D-QOCCCs to map 1-qubit to 13-qubits, with the ability to correct 2 errors.    
    }}
    \label{fig:6}
\end{figure}

Table~\ref{tab:1-13 qubits} illustrates the simulation results for 2D-QOCCCs involving both QO and GT along with their corresponding count frequencies. Notably, the outcomes $1111000100011$ and $1110010001100$ stand out with the highest repetition counts, both at seven, indicating their frequent recurrence. Conversely, the remaining outcomes have lower counts, implying lower accuracy. The recurring instances of these outcomes suggest the successful correction of two errors.

 Figure~\ref{fig:5}(a) shows the initial 1-qubit diagram, where the application of the $X$ gate confirms the identification of two errors for correction, since it has only one qubit. Next, Figure~\ref{fig:5}(b) shows the final 13-qubit diagram where the $X$-gate is used along with parity and auxiliary qubits to correct the identified errors.

Figure~\ref{fig:6} shows the plots for QO and GT over 2D-QOCCCs, transforming 1-qubit into 13-qubits with the ability to correct two errors. Figure~\ref{fig:6}(a) shows results characterized by the presence of the binary digit $'1'$, while Figure~\ref{fig:6}(b) shows results with $'0'$ in the binary code. As indicated in Table~\ref{tab:1-13 qubits}, the presence of only two outcomes with the highest frequencies, represented by taller bars in these plots, further confirms the correction of 2 errors.

\item [\textbf{$C_4$:}] Finally, when mapping 1-qubit to 29-qubits, we can correct up to 5 errors.
\begin{table}[!ht]
\caption{
Simulation results of 2D-QOCCCs for both QO and GT to map 1-qubit to 29-qubits with 5 error corrections.}
    \label{tab:1-29 table}
    \begin{tabular}{c|c}
       Outcome (qubits)  & Counts \\ \hline
        00010001000101100101101010011 & 1\\
        00001001000011101000100111000&3\\
        11110001000110100001011110011&7\\
        11100100011011111010001111110&7\\
         00000111000010110110111011001&5\\
         00001001000011101000100111000&3\\
         10010001000101110001110101111&7\\
         1100100011011111010001111110&7\\
         01000000100000001000010100000&4\\
         10110101110101111111111111010&7\\ \hline
    \end{tabular}
 
\end{table}


\begin{figure}[!ht]
     \subfigure[1-qubits]{\includegraphics[scale=0.20]{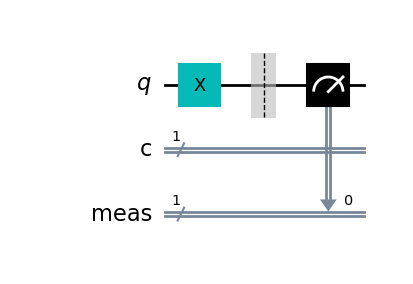}}
     \subfigure[29-qubits]{\includegraphics[scale=0.12]{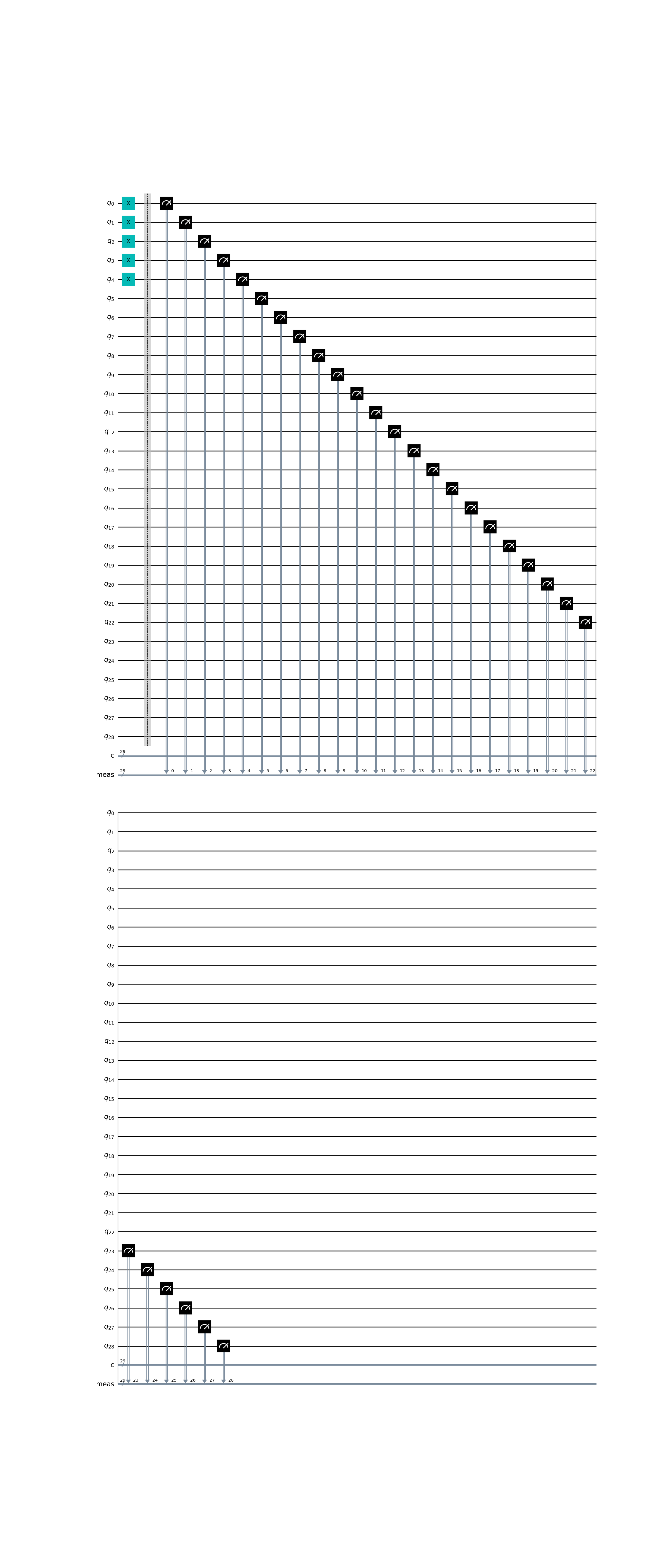}}
    \caption{Initial and final quantum states corresponding to 1-qubit and 29-qubits, with identified errors.}
    \label{fig:7}
\end{figure}

\begin{figure}[!ht]
     \subfigure[QO (2D-QOCCCs) for 1-qubit to 29-qubits]{ \includegraphics[scale=0.25]{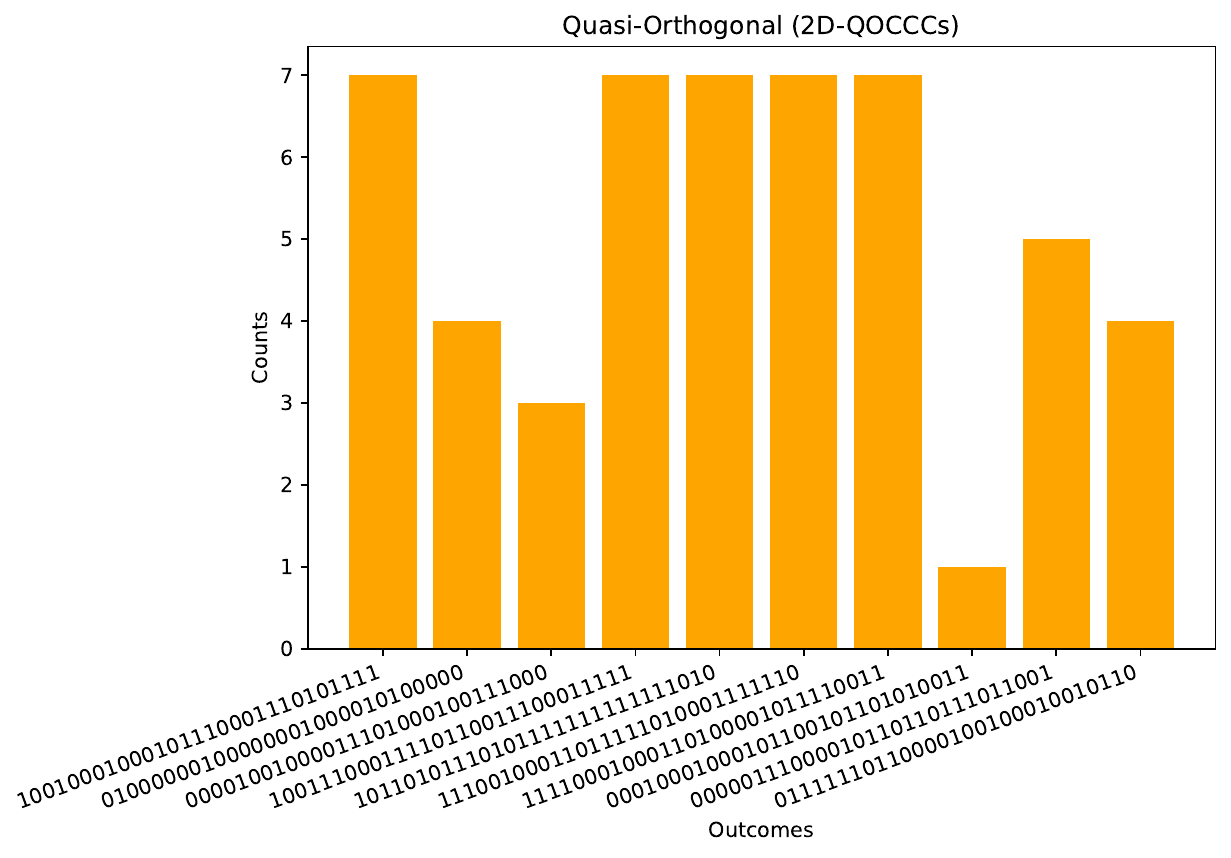}}
     \subfigure[GT (2D-QOCCCs) for 1-qubit to 29-qubits] {\includegraphics[scale=0.25]{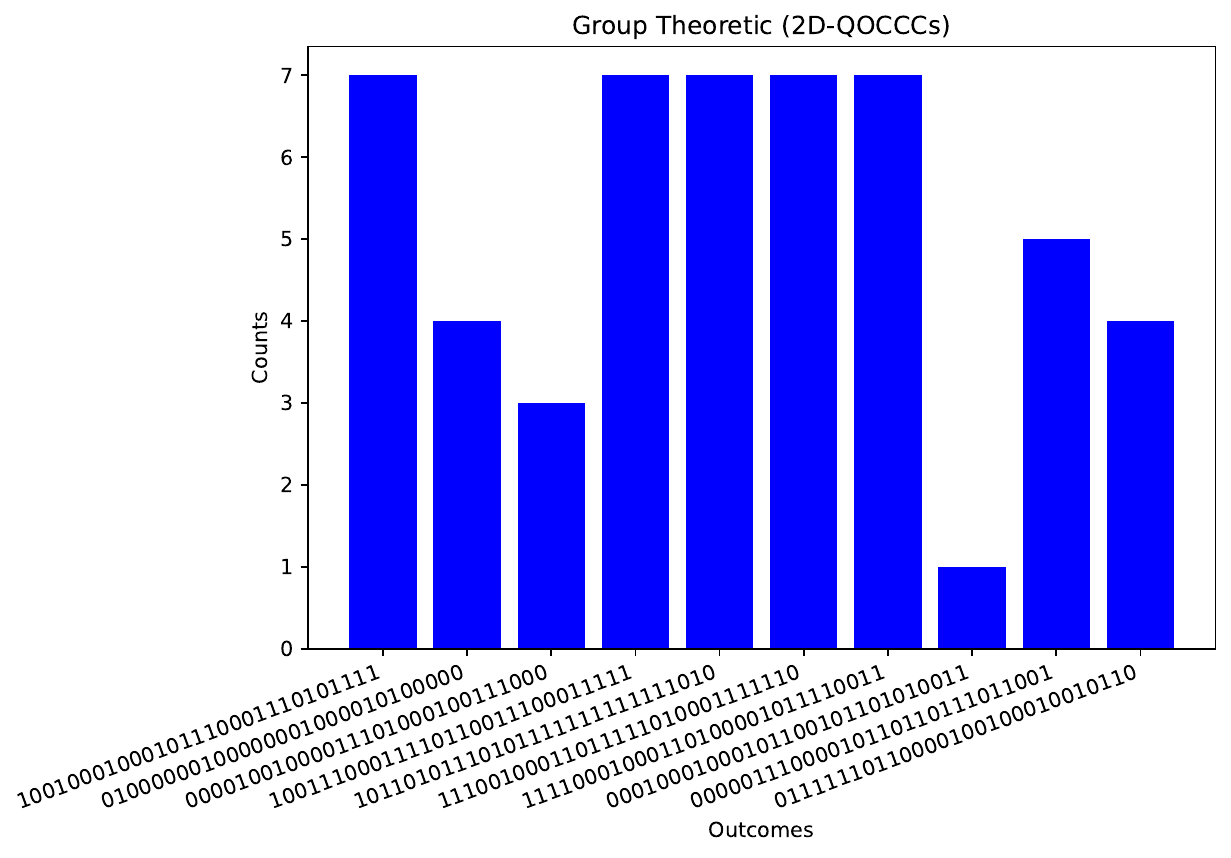}}
    \caption{ 
    Plots of GT and QO for 2D-QOCCCs to map 1-qubit to 29-qubits, with the ability to correct 5 errors. 
    }
    \label{fig:8}
\end{figure}

Table~\ref{tab:1-29 table} shows the simulation results of 2D-QOCCCs under both QO and GT along with their respective counts. The five results listed below\\

$11110001000110100001011110011$\\
$11100100011011111010001111110$\\ 
$10010001000101110001110101111$\\
$1100100011011111010001111110$\\ 
$10110101110101111111111111010$\\

have the highest counts (7), indicating their frequent recurrence.  Conversely, other outcomes have lower counts, indicating less accuracy in correcting errors. Repeating these results five times means that five errors have been successfully corrected.

In Figure~\ref{fig:7}(a), which shows the initial 1-qubit circuit diagram, a single application of the $X$-gate detects an error. Four subsequent applications are applied to the final qubits, as shown in Figure~\ref{fig:7}(b). This final diagram contains 29-qubits, integrating parity, and auxiliary qubits to correct the identified errors.

Figure~\ref{fig:8} illustrates the plots for QC and GT over 2D-QOCCCs, transforming 1-qubit into 29-qubits with the ability to correct five errors. Figure~\ref{fig:8}(a) shows results characterized by the presence of the binary digit $"1"$, while Figure~\ref{fig:8}(b) shows results with $"0"$ in the binary code. As observed in Table~\ref{tab:1-29 table}, the presence of five outcomes with the highest frequencies, represented by taller bars in these plots, further confirms the correction of five errors.
\end{enumerate}
In the general exploration of plots, specifically in Figures~\ref{fig:2}, \ref{fig:4}, \ref{fig:6}, \ref{fig:8}(a), we present plots for 2D-QOCCCs with a focus on the results arising from the 2D-QOCCCs characterized by the presence of the binary digit $"1"$ in their representation. Similar to the methodology used to construct the GT plots, a systematic approach was used to randomly select quantum states and meticulously record their respective frequencies of occurrence.
Of particular note is the central role of QO in the constitution of these codes, which delineates the nearly orthogonal relationships between different quantum states. This feature assumes a critical role in the structural design and operational functionality of 2D-QOCCCs, effectively illustrating the nearly orthogonal connections that prevail between different quantum states. In contrast to the results observed in the GT plots, the distribution of results in the QO plots is influenced by the nearly orthogonal properties inherent in QO codes.
These unique QO properties contribute to a distinct pattern in the distribution of results, highlighting the inherent uniqueness of these codes in generating quantum states with near-orthogonal connections. This particular property has significant value in several applications, with a particular emphasis on communication and coding theory.
\parskip 0.1in
The plots in Figure~\ref{fig:2}, \ref{fig:4}, \ref{fig:6}, \ref{fig:8}(b) show the application of GT in the domain of 2D-QOCCCs. The bars represent results with $'0'$ in the binary code, which may show some symmetry. The plots describe the frequency distribution of each quantum state depending on its group-theoretic properties, with emphasis on the symmetry and interrelationships that exist between quantum states according to the properties of GT. The selected quantum states showed different results, each with its count. This variety suggests that 2D-QOCCCs are designed to give different results based on the properties of GT. This observation is related to the main ideas of 2D-QOCCCs, and these ideas are based on GT.
The simulation provides valuable insights into the distinct properties and diversity of outcomes associated with different binary representations. These visualizations deepen our understanding of the behavior of 2D-QOCCCs and their potential applications in quantum information processing and communication. While there are some differences between GT and QO plots, each of which exhibits unique features such as organized patterns in GT properties and near-orthogonality in QC codes, their comparable performance  based on the number of quantum states is evident. Further exploration of different sets of quantum states may provide additional details about the performance and characteristics of these codes. 
\parskip 0.1in

In addition to comparing the number of quantum states between the GT and QO plots, the simulation sheds light on another important concept: probability amplitudes. Probability amplitudes, a fundamental aspect of quantum mechanics that represents the probability of a quantum system being in a particular state, can be inferred from the height of the bars in the plots. These amplitudes are central to quantum mechanics, with the square of the magnitude proportional to the probability of observing a particular outcome.
In the GT plots, the observation suggests a pattern influenced by group-theoretic properties described by GT and linear algebra, possibly leading to symmetries and structured probability amplitudes. Conversely, the QO plots suggest different patterns of probability amplitudes influenced by the near-orthogonal nature of the QC codes. The analysis of probability amplitudes in conjunction with counts provides a comprehensive understanding of the behavior of quantum states, offering insights into both frequency and probability within the framework of 2D-QOCCCs.

\subsection{AQECCs case}\label{section 4.2}

We simulate and compare AQECCs constructed using QC codes and GT properties for equation \ref{17}. The experimental results are consistent with the properties of GT and QC, especially for quantum states with $'0'$ and $'1'$ in their binary codes. The application of $H$-gates to logical qubits forms the basis of both codes, introducing quantum superposition and allowing the exploration of quantum states.
\parskip 0.1in
A mathematical formalism is then used to integrate these quantum states and examine their frequency and impact in detail. Bar graphs are then generated to visually represent the results. Given the large number of results from the simulation, a random sample is selected for comparison and analysis. A comprehensive analysis is then performed to evaluate the performance of transforming qubits with AQECCs using QC codes and the GT approach.
The code examples provided demonstrate the encoding of logical qubits into physical qubits, accompanied by error correction using QC structures and $H$-gates. 
\parskip 0.1in
Consider the following four cases of section \ref{section 3} in the configurations where $N$-qubits map into $M$-qubits and have the ability to correct $P$ errors:
\begin{enumerate}
   \item[\textbf{$C_1$:}]For mapping 3-qubits to 8-qubits, we can correct 1 error.

    \begin{table}[!ht]
    \centering
         \caption{Simulation results of AQECCs for both QO and GT to map 3-qubits to 8-qubits with one error correction.
         }
         \label{Table:5}
    \begin{tabular}{|c|c|c|} 
    \hline 
   \multicolumn{1}{|c|}{ Outcomes} &
   \multicolumn{2}{|c|}{Counts}   \\ \hline 
\textbf{} &  QC &  GT \\ \hline
    00000100& 6 & 7\\
    00000001 & 8 & 13\\
    00000101 & 5 & 11\\
    00000010&5&11\\
    00000000&18&11\\
    00000111&11&9\\
    00000011&16&10\\
    00000110&11&8\\
    \hline
    \end{tabular}
    \end{table}

\begin{figure}[!ht]
    \subfigure[3-qubits]{\includegraphics[scale=0.25]
    	{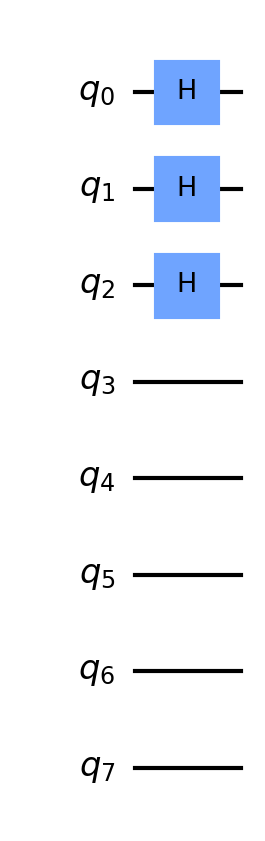}}
    \hspace{0.5cm} 
    \subfigure[8-qubits]{\includegraphics[scale=0.25]
    	{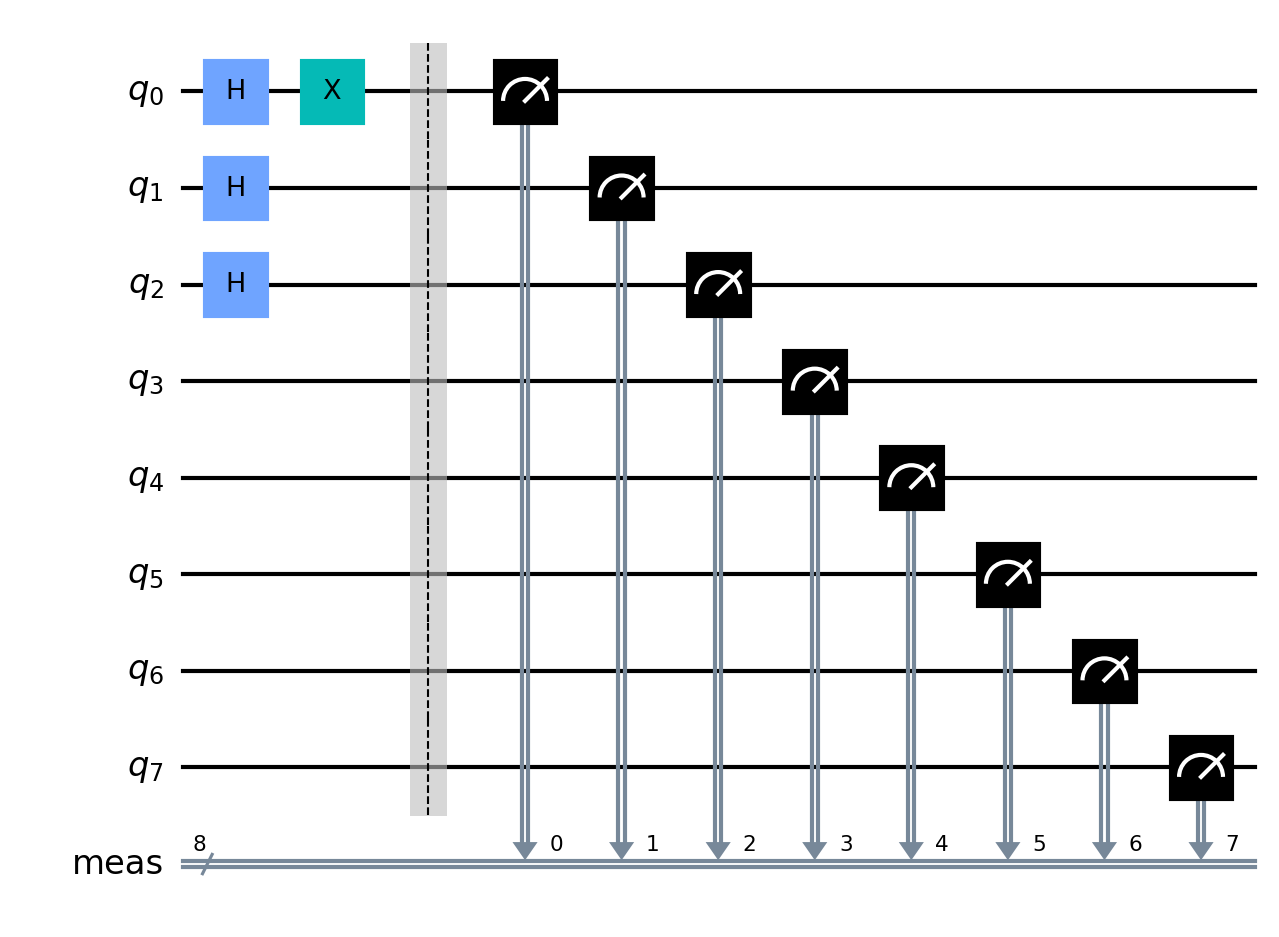}}
    \caption{ 
    Initial and final quantum states with the application of $H$-gate to map 3-qubits to 8-qubits,  with identified errors.}
    \label{fig:99}
\end{figure}

\begin{figure}[!ht]
    \centering
     \subfigure[QC (AQECCs) for 3-qubits to 8-qubits]{ \includegraphics[scale=0.30]{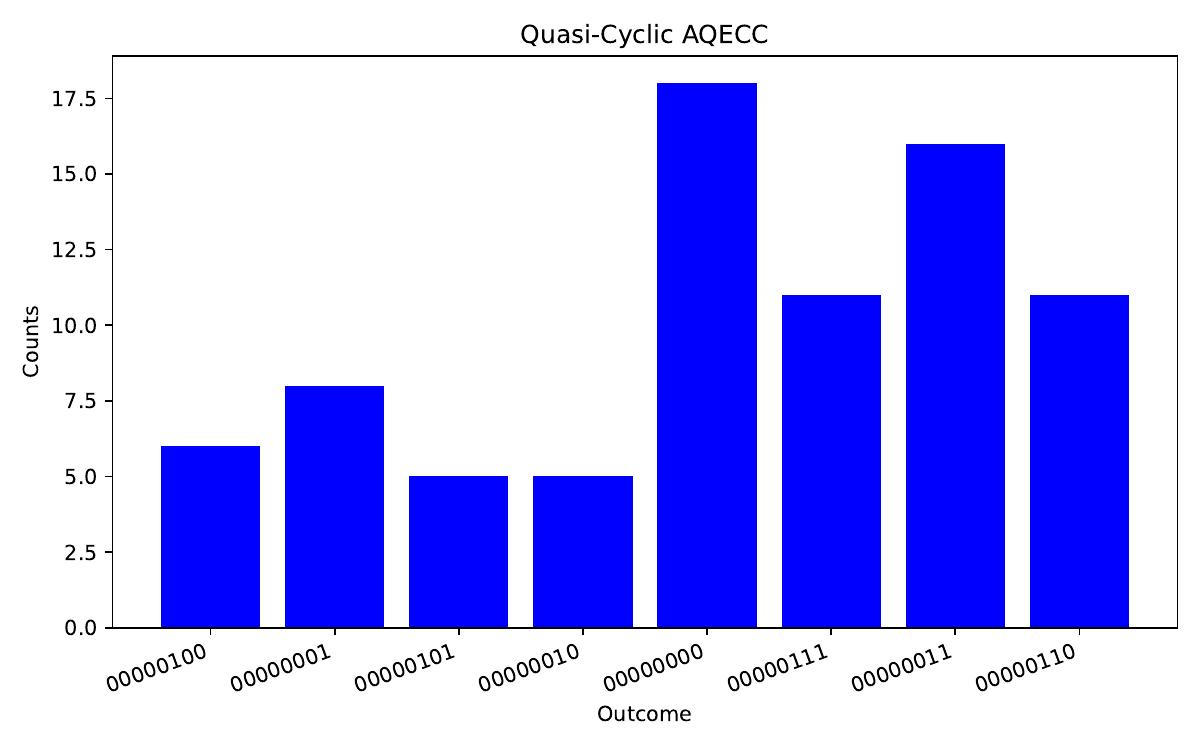}}
     \subfigure[GT (AQECCs) for 3-qubits to 8-qubits] {\includegraphics[scale=0.30]{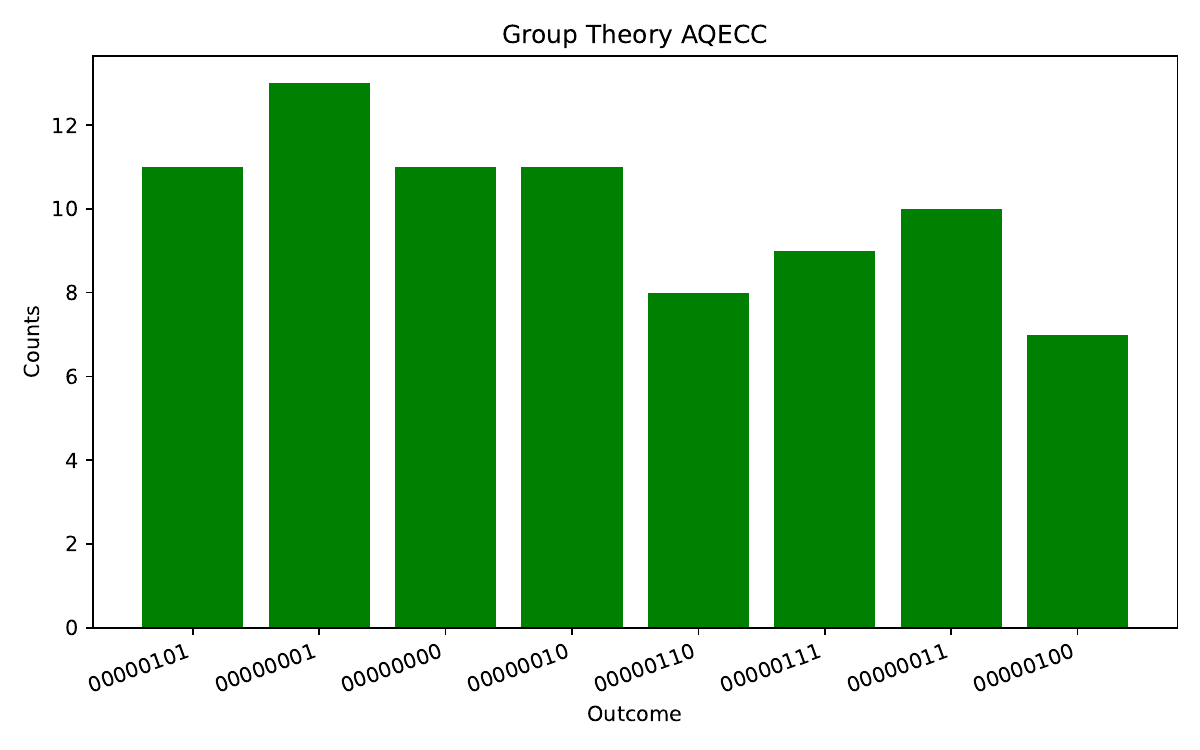}}
    \caption{ Plots of GT and QO for AQECCs to map 3-qubits to 8-qubits, with the ability to correct one error. 
    }
    \label{fig:10}
\end{figure}

\begin{figure}[!ht]
    \centering
   
    \includegraphics[scale=0.30]{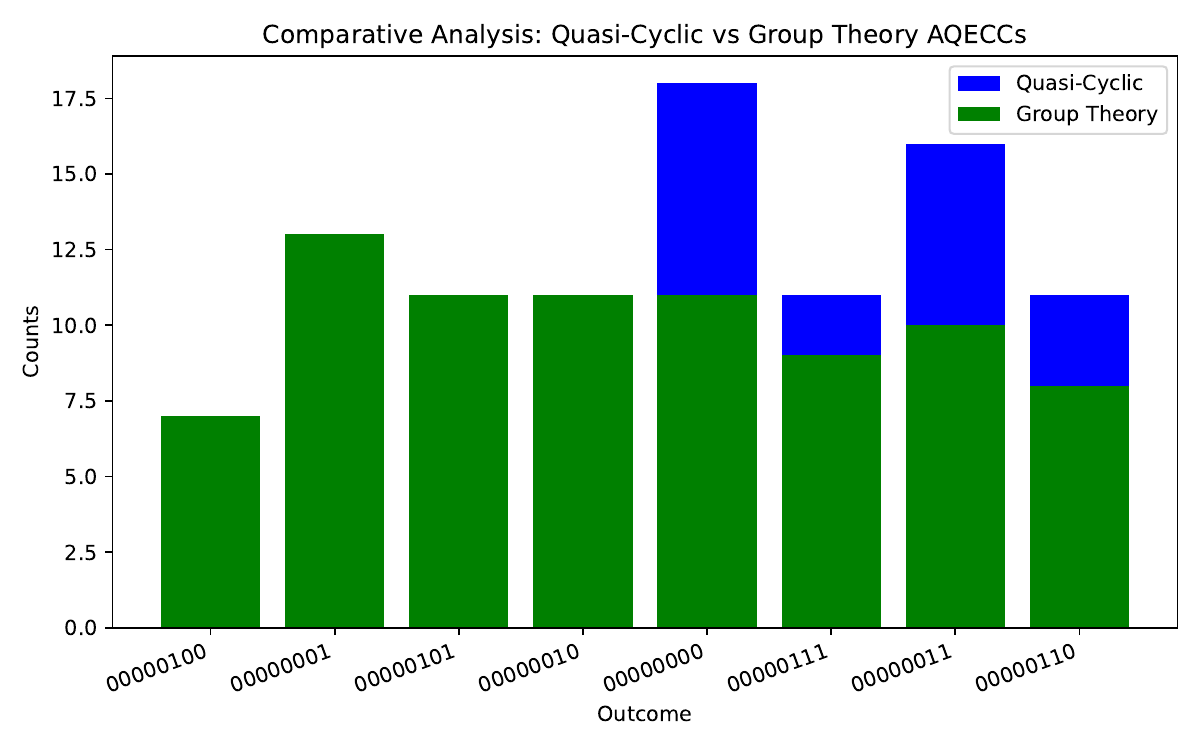}
     \caption{Comparative analysis between GT and QC properties when transforming  3-qubits to 8-qubits under AQECCs.}
    \label{fig:11}
\end{figure}

Table~\ref{Table:5} shows the simulation results for AQECCs using the QC and GT approaches. Notably, outcomes $00000000$ and $00000001$ have the highest frequencies in QC and GT, respectively, indicating their frequent occurrence in the AQECC simulation. These counts indicate the likelihood of certain measurement results, with the high-frequency results indicating successful single-error correction.

Figure~\ref{fig:99}(a) shows a schematic of the application of $H$-gates to induce superposition in logical qubits. The subsequent measurement collapses the superposition, revealing results with different probabilities, while Figure~\ref{fig:99}(b) shows the final 8-qubit scheme with the application of the $X$-gate to identify the error, confirming that an error has been identified for correction. 

Figure~\ref{fig:10} illustrates the plots showing QC and GT under AQECCs for transforming 3-qubits to 8-qubits with the ability to correct an error. In Figure~\ref{fig:10}(a), the blue bar plot shows the results of AQECCs characterized by QC properties, where the distribution of counts elucidates the probabilities associated with different measurement results with the high performance of error correction.
Figure~\ref{fig:10}(b) then shows the outcomes of GT under AQECC, presented in a green bar graph. The counts for each result provide insight into the effectiveness and precision of the ECCs, with the distribution influenced by quantum interference and entanglement introduced by $H$-gates closer. As observed in Table~\ref{Table:5}, the prevalence of only one outcome with high frequency for each number confirms the correction of an error. However, it is crucial to note that the error correction capability varies with different parameters, such as code distance and minimum distance.

Figure~\ref{fig:11} shows a comparative analysis between GT and QC properties regarding the transformation from 3-qubits to 8-qubits under AQECCs. In particular, the QC properties show superior performance in correcting a single error, as shown in the tallest bar compared to the bars representing the GT properties.

\item [\textbf{$C_2$:}] When mapping 4-qubits to 10-qubits, we still correct 1 error.

\begin{table}[!ht]
\centering
   \captionof{table}{Simulation results of AQECCs for both QO and GT to map 4-qubits to 10-qubits with one error correction.}
   \label{Table:6}
   \begin{tabular}{|c|c|c|} 
    \hline 
   \multicolumn{1}{|c|}{ Outcomes} &
   \multicolumn{2}{|c|}{Counts}   \\ \hline 
\textbf{} &  QC &  GT \\ \hline
    0000000010& 2 & 8\\ 
    0000001101 & 4&9\\
    0000001111 & 5 &5\\
    0000001100 & 9 &10\\
    0000000011&6&4\\
    0000000111&4&3\\
    0000001011&5&2\\
    0000001001&4&6\\
    0000000101&4&2\\
    0000001110&5&5\\
    0000001000&5&3\\
    0000000110&6&9\\
    0000001010&8&3\\
    0000000000&4&5\\
    0000000100&5&5\\
    0000000001&4&2\\
    
    \hline
  \end{tabular}

\end{table}

\begin{figure}[!ht]
    \subfigure[4-qubits]{\includegraphics[scale=0.25]
    	{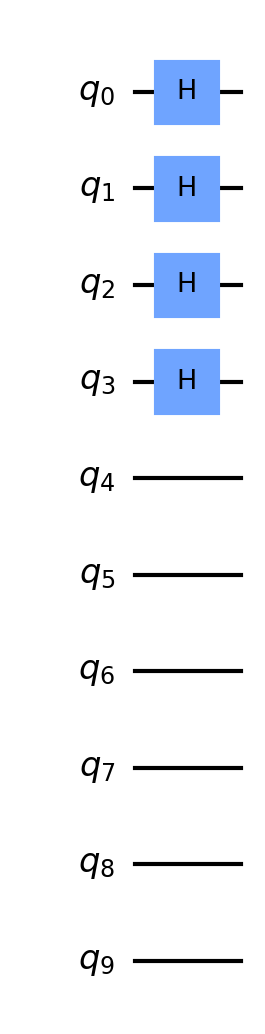}}
    \hspace{0.5cm} 
    \subfigure[10-qubits]{\includegraphics[scale=0.25]
    	{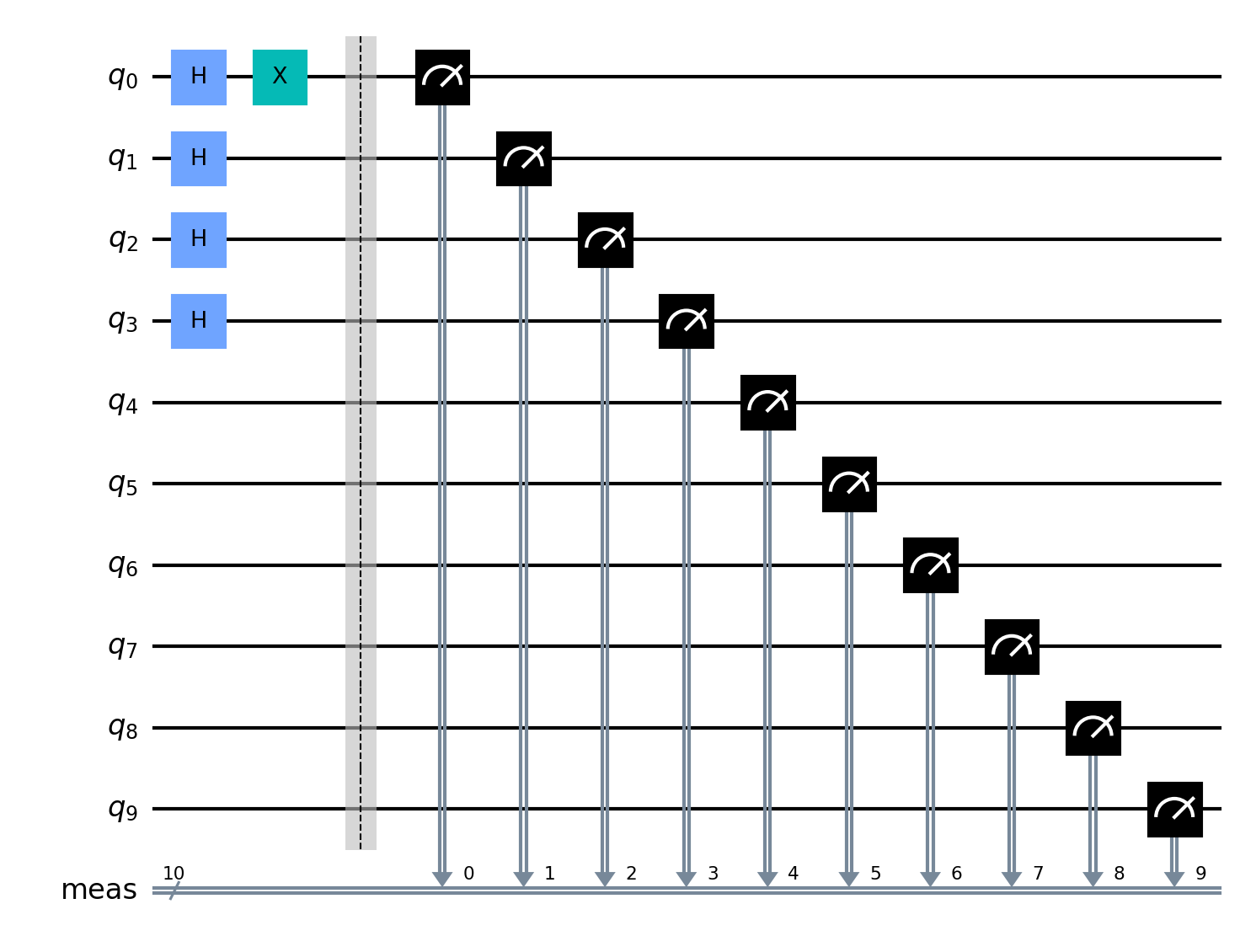}}
    \caption{Initial and final quantum states with $H$ gate applied to 4-qubits to get 10-qubits, and identified errors.}
    \label{fig:12}
\end{figure}

\begin{figure}[!ht]
    \centering
     \subfigure[QC (AQECCs) for 4-qubits to 10-qubits]{ \includegraphics[scale=0.30]{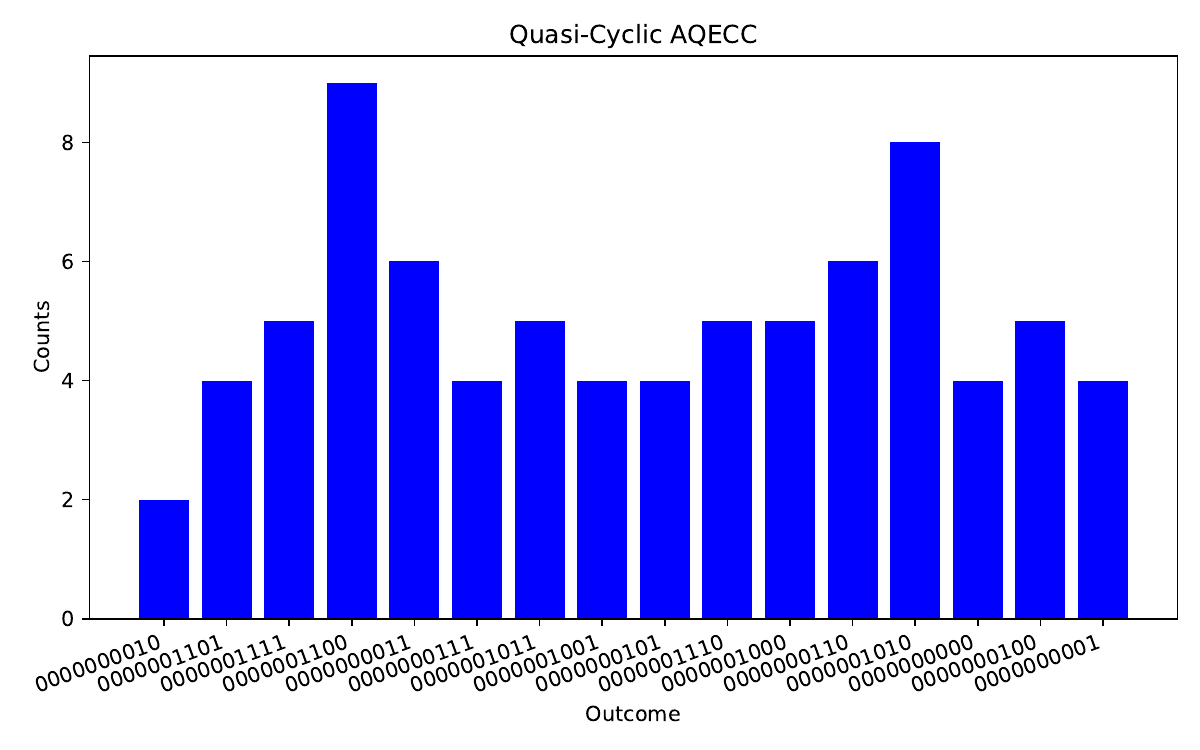}}
     \subfigure[GT (AQECCs) for 4-qubits to 10-qubits] {\includegraphics[scale=0.30]{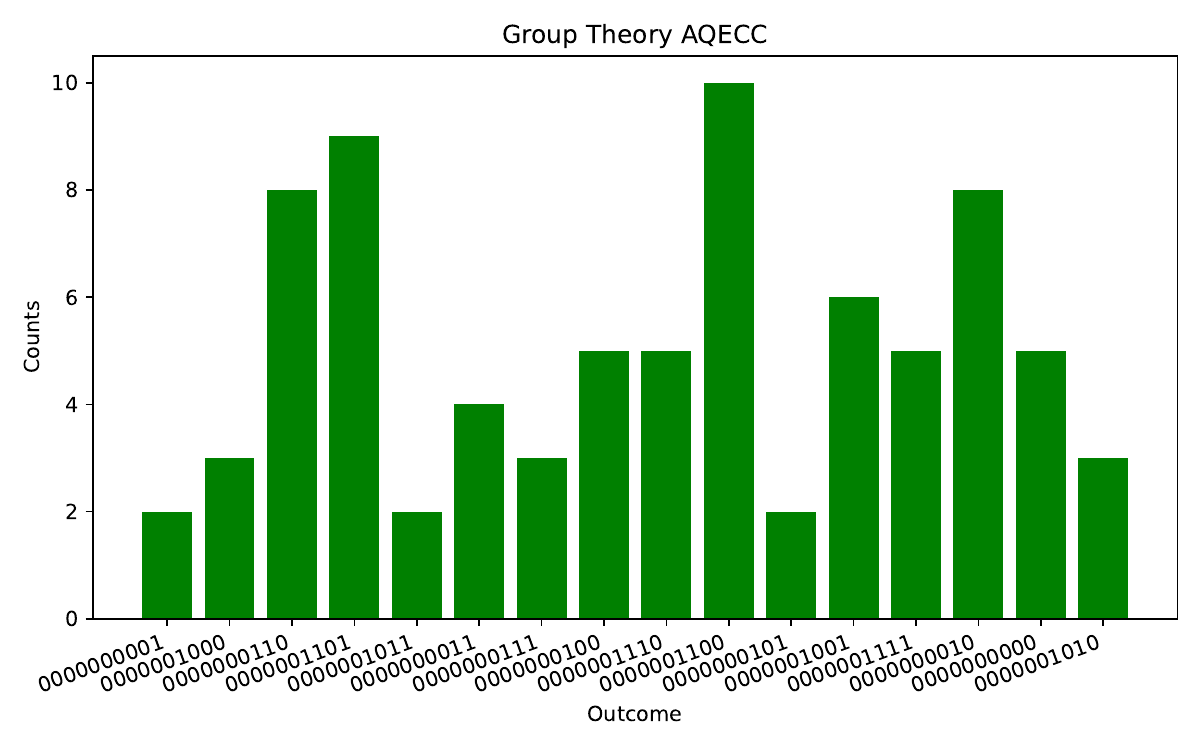}}
    \caption{
Plots of GT and QO for AQECCs to map 4-qubit to 10-qubits, with ability to correct one error.    
    }
    \label{fig:13}
\end{figure}

\begin{figure}[!ht]
    \centering
    
    \includegraphics[scale=0.30]{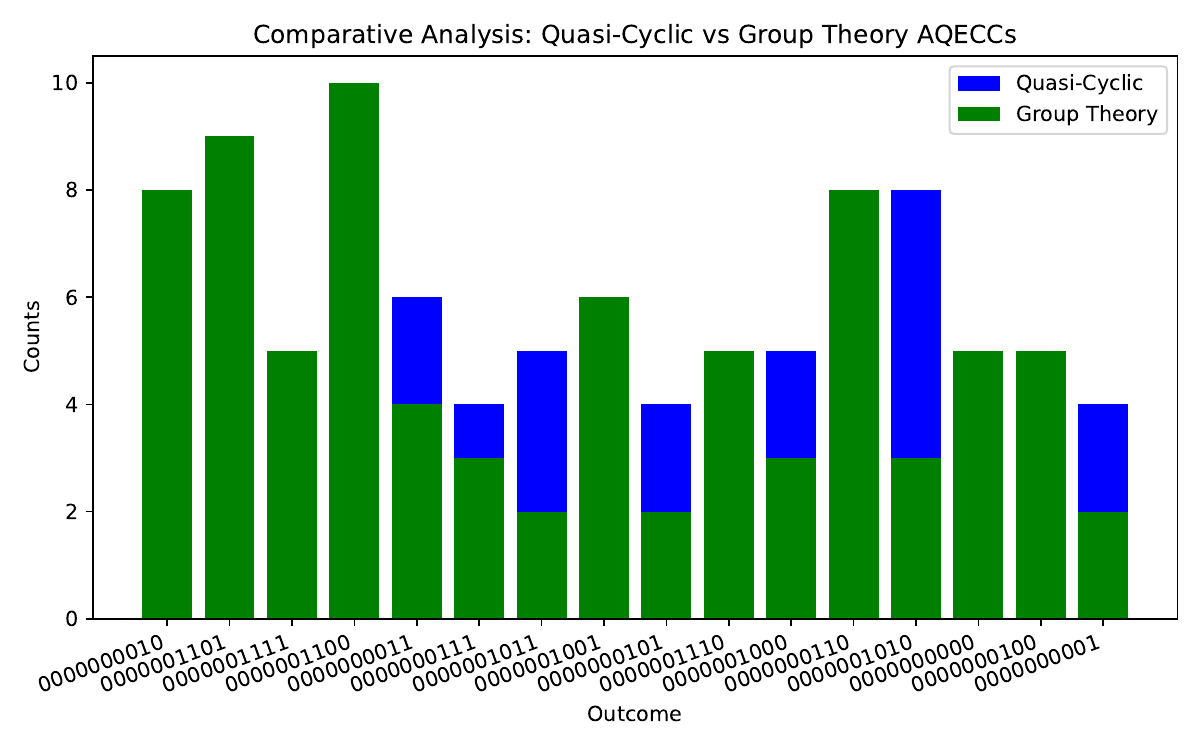}
    \caption{Comparative analysis between GT and QC properties when transforming  4-qubits to 10-qubits under AQECCs.}
    \label{fig:14}
\end{figure}

Table~\ref{Table:6}, shows the simulation results for AQECCs using the both QC and GT approaches. In particular, the result $0000001100$ reaches the highest frequencies of $9$ and $10$ in QC and GT respectively. This indicates its frequent occurrence in the process of transforming 4-qubits into 10-qubits.  In contrast, other outcomes have lower counts, suggesting less accuracy in error correction. The consistent occurrence of this result underscores the successful correction of a single error.

Figure~\ref{fig:12}(a) shows a circuit illustrating the use of $H$-gates to induce superposition in 4-qubits. The measurement shows superposition with different probabilities. Figure~\ref{fig:12}(b) shows the circuit for physical qubits using $X$-gates to identify and confirm the correction of a single error.  

Figure~\ref{fig:13} shows plots of QC and GT under AQECCs for transforming a smaller number of qubits into a larger set with the ability to correct a single error. Figure~\ref{fig:13}(a) shows outcomes characterized by QC properties, providing insight into the probabilities associated with different outcomes with precision. Figure~\ref{fig:13}(b) shows GT results, providing further insight into the effectiveness and high performance of the ECCs.
As noted in Table~\ref{Table:6}, the widespread occurrence of a single outcome with different high frequencies in both figures confirms the correction of an error. However, the frequency varies, indicating a different ability to correct errors based on parameters such as code distance and minimum distance.

Figure~\ref{fig:14} shows a comparative analysis between GT and QC properties in the form of grouped stacked bars for the transformation of 4-qubits to 10-qubits under AQECCs. However, the GT properties show strong performance in correcting a single error, which is evident in the tallest bar in contrast to the bars representing the QC properties.

\item [\textbf{$C_3$:}]When mapping 1-qubit to 13-qubits, we can correct 2 errors. 
\begin{table}[!ht]
    \centering
  \captionof{table}{
  Simulation results of AQECCs for both QO
and GT to map 1-qubit to 13-qubits with 2 error
corrections.}
   \label{Table:7}
   \begin{tabular}{|c|c|c|} 
    \hline 
   \multicolumn{1}{|c|}{ Outcomes} &
   \multicolumn{2}{|c|}{Counts}   \\ \hline 
\textbf{} &  QC &  GT \\ \hline
    0000000000000&36&32\\
    0000000000001&44&48\\
    \hline
    \end{tabular}
\end{table}

\begin{figure}[!ht]
    \subfigure[1-qubit]{\includegraphics[scale=0.20]
    	{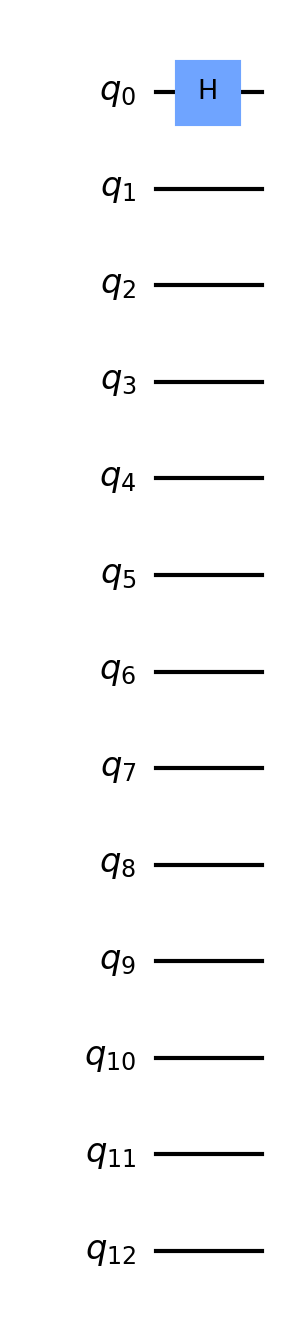}}
    \hspace{0.5cm} 
    \subfigure[13-qubits]{\includegraphics[scale=0.22]
    	{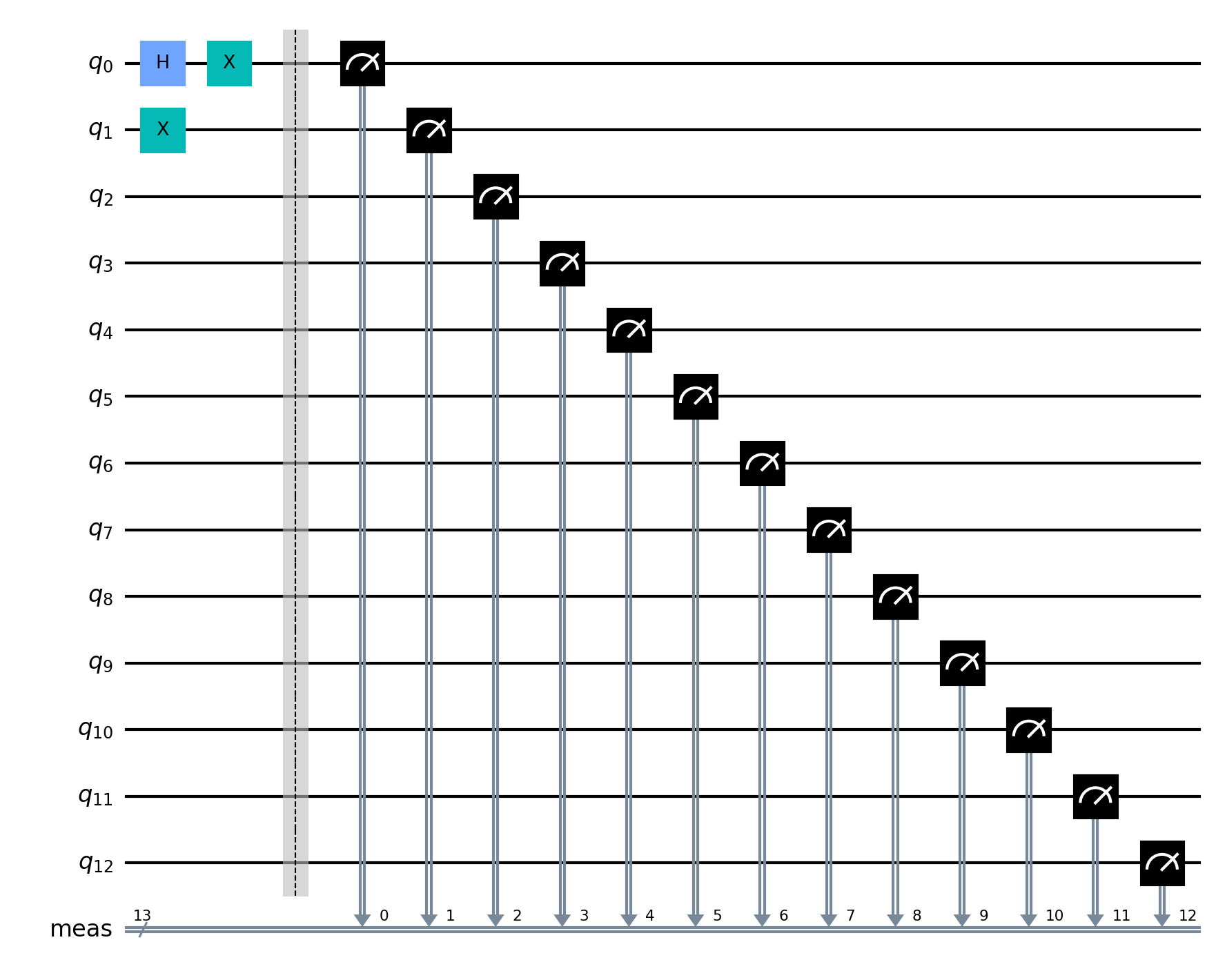}}
    \caption{Initial and final quantum states with $H$ gate applied to 1-qubit to get 13-qubits, and identified errors.
 }
    \label{fig:15}
\end{figure}

\begin{figure}[!ht]
    \centering
     \subfigure[QC (AQECCs) for 1-qubit to 13-qubits]{ \includegraphics[scale=0.30]{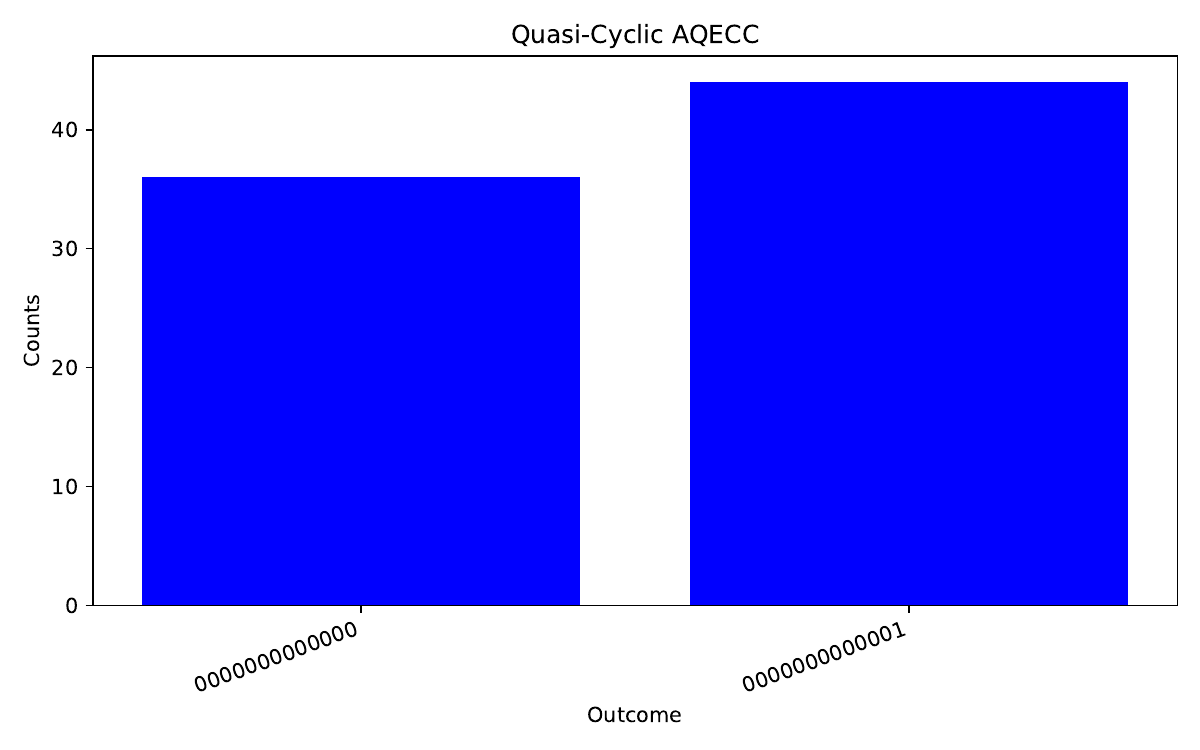}}
     \subfigure[GT (AQECCs) for 1-qubit to 13-qubits] {\includegraphics[scale=0.30]{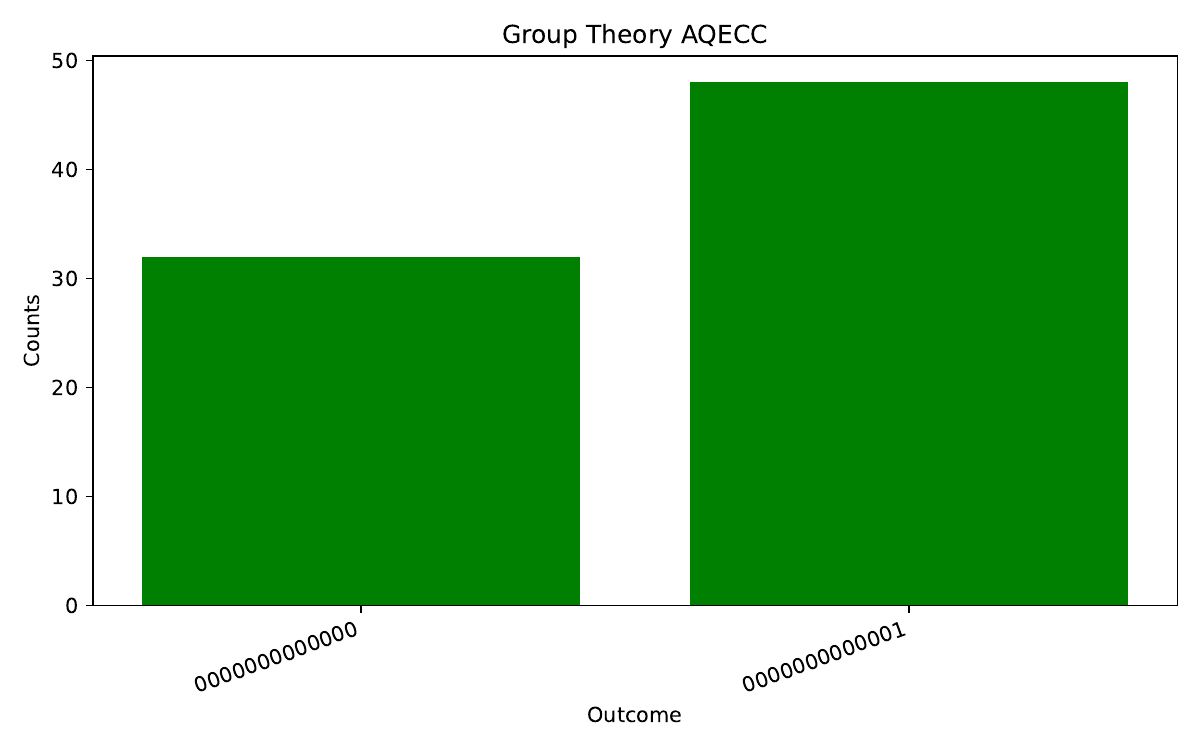}}
    \caption{Plots of GT and QO for AQECCs to map
1-qubit to 13-qubits, with the ability to correct 2 errors.
 }
    \label{fig:16}
\end{figure}

\begin{figure}[!ht]
    \centering
    
    \includegraphics[scale=0.30]{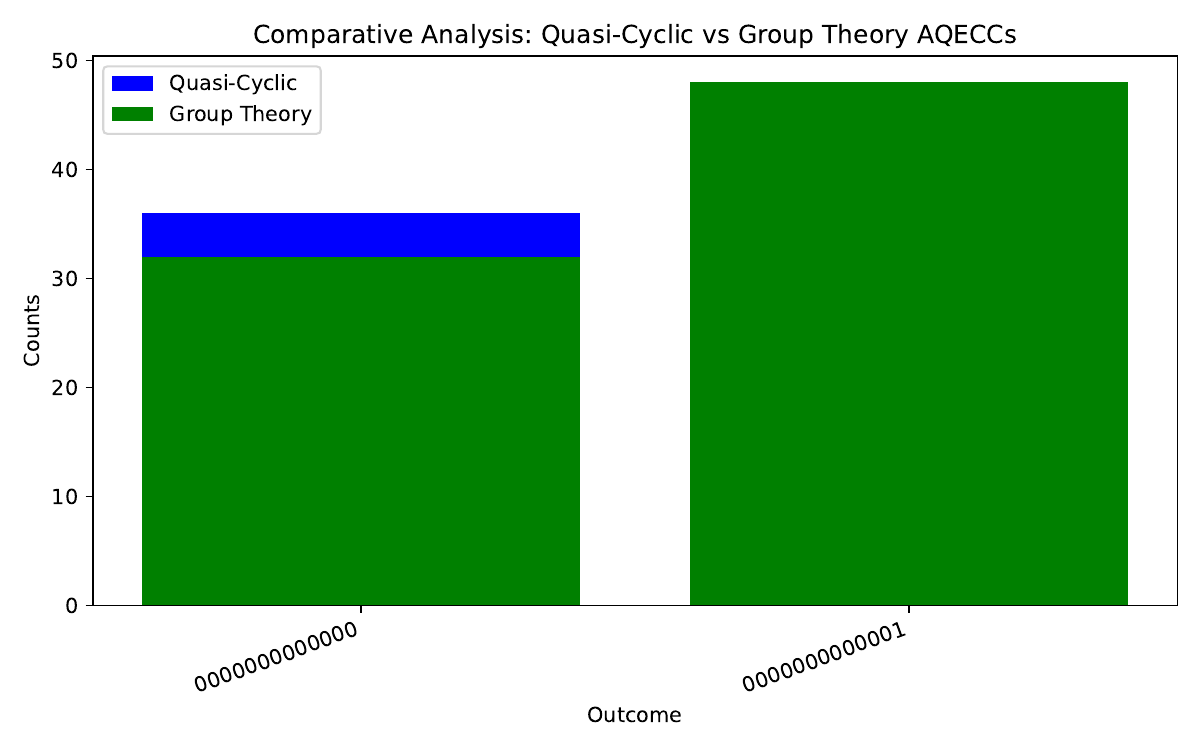}
    \caption{Comparative analysis between GT and QC
properties when transforming 1-qubit to 13-qubits under AQECCs.
   }
    \label{fig:17}
\end{figure}

Table~\ref{Table:7} shows the simulation results of AQECCs for QC and GT, aiming to map 1-qubit into 13-qubits while correcting 2 errors. However, the table does not show many counts, reflecting the inherent characteristics of the simulated QECCs. In this code, a small number of qubits are mapped to larger ones. Since the quantum circuits are specifically designed to correct only 2 errors, the counts for outcomes with more errors are significantly lower.

Figure~\ref{fig:15} shows the exploration of the initial and final quantum states, the application of the $H$-gate, and the identification of errors with the $X$-gate for 1-qubit in 13-qubits. Specifically, in Figure~\ref{fig:15}(a), the quantum circuit diagram shows the application of the $H$-gate to a single qubit. In addition, Figure~\ref{fig:15}(b) shows the circuit diagram incorporating the Pauli $X$-gate to identify and confirm the correction of two errors in this transformation.

 Figure~\ref{fig:16} shows the properties of GT and QC under AQECCs for the process of mapping 1-qubit to 13-qubits, demonstrating the ability to correct 2 errors. In Figure~\ref{fig:16}(a), the plot of QC under AQECCs shows two results, indicating error correction capability and consistency in the process. Figure~\ref{fig:16}(b) also shows high efficiency and consistency in correcting two errors using the GT properties, with reduced counts for outcomes with more errors. Figure~\ref{fig:17} shows a grouped stacked bar chart comparing QC and GT properties in this process, both of which indicate success. In particular, GT shows high performance on the $0000000000001$ outcome, represented by the tallest bar in the graph compared to the others.

\item [\textbf{$C_4$:}] Finally, when mapping 1-qubit to 29-qubits, we can correct up to 5 errors.

\begin{table}[!ht]
  \centering
  \captionof{table}{
  Simulation results of AQECCs for both QO and
GT to map 1-qubit to 29-qubits with 5 error corrections.
  }
   \label{Table:8}
   \begin{tabular}{|c|c|c|} 
    \hline 
   \multicolumn{1}{|c|}{ Outcomes} &
   \multicolumn{2}{|c|}{Counts}   \\ \hline 
\textbf{} &  QC &  GT \\ \hline
   00000000000000000000000000000&38&48\\
   00000000000000000000000000001&42&32\\
    
    \hline
  \end{tabular}
\end{table}

\begin{figure}[!ht]
    \centering
    \subfigure[1-qubit]{\includegraphics[scale=0.10]
    	{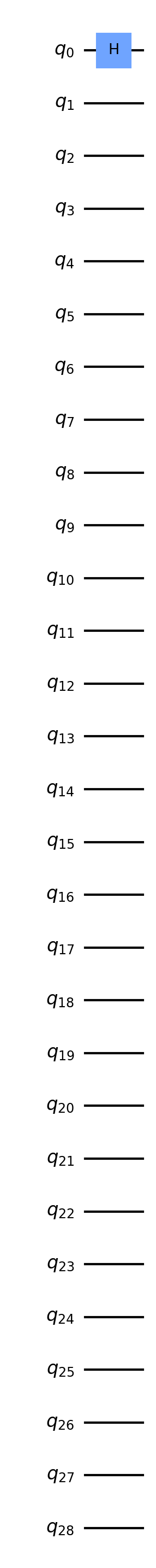}}
    \hspace{0.5cm} 
    \subfigure[29-qubits]{\includegraphics[scale=0.15]
    	{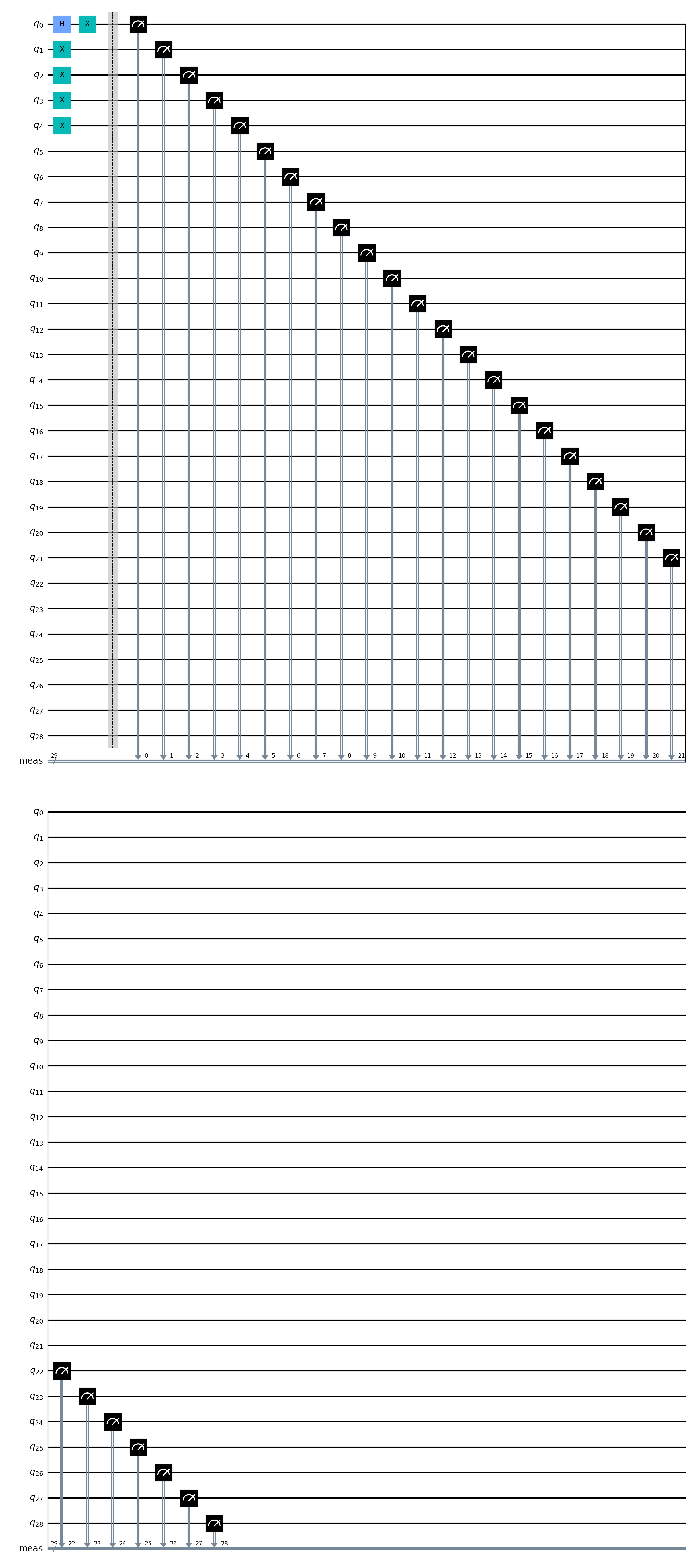}}
    \caption{ Investigation of the initial and final quantum states through the application of the $H$-gate while identifying 5 errors with the $X$-gate in the context of mapping 1-qubit into 29-qubits.}
    \label{fig:18}
\end{figure}

\begin{figure}[!ht]
    \centering
     \subfigure[QC (AQECCs) for 1-qubit to 29-qubits]{ \includegraphics[scale=0.30]{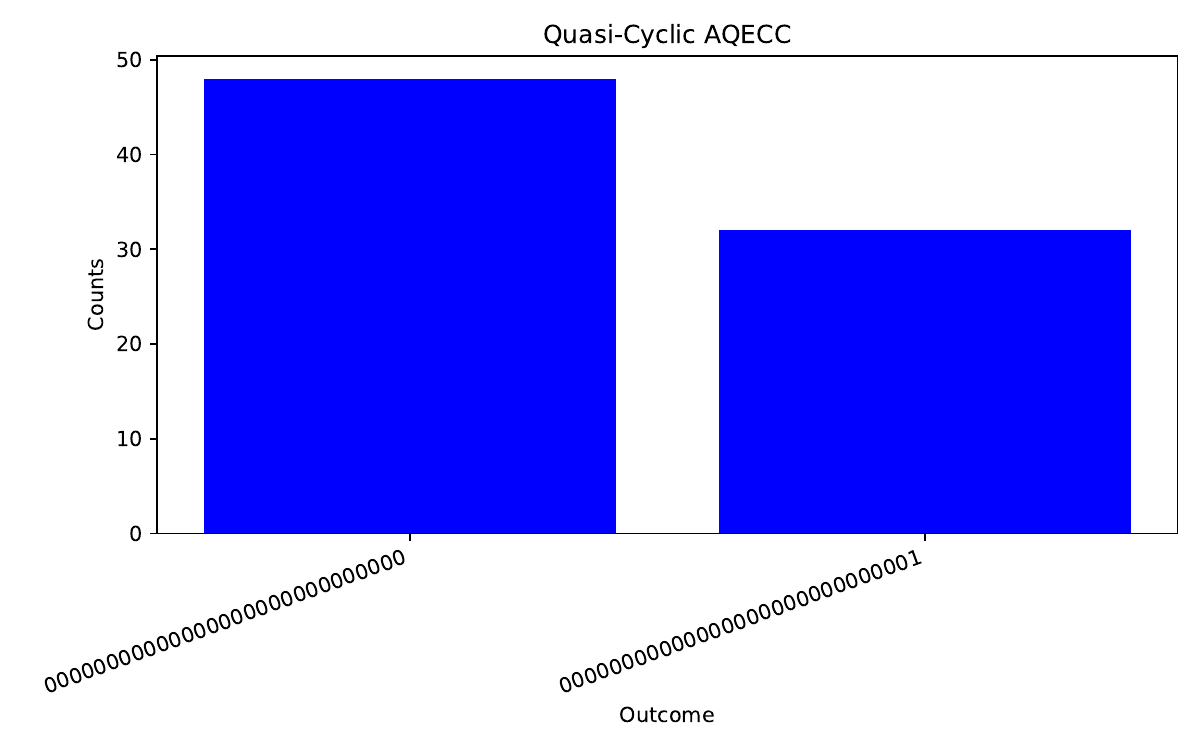}}
     \subfigure[GT (AQECCs) for 1-qubit to 29-qubits.] {\includegraphics[scale=0.30]{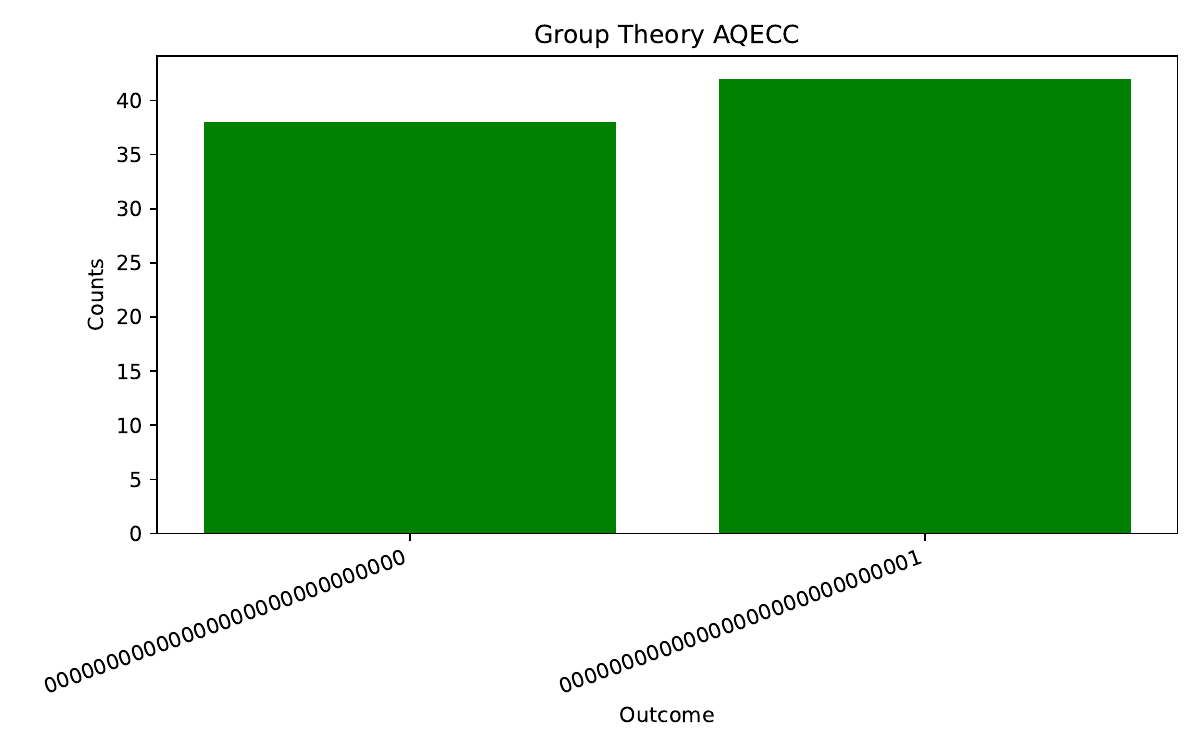}}
    \caption{ Plots of GT and QO for AQECCs to map
1-qubit to 29-qubits, with the capability of correcting 5 errors.
}
    \label{fig:19}
\end{figure}

\begin{figure}[!ht]
    \centering
    
    \includegraphics[scale=0.30]{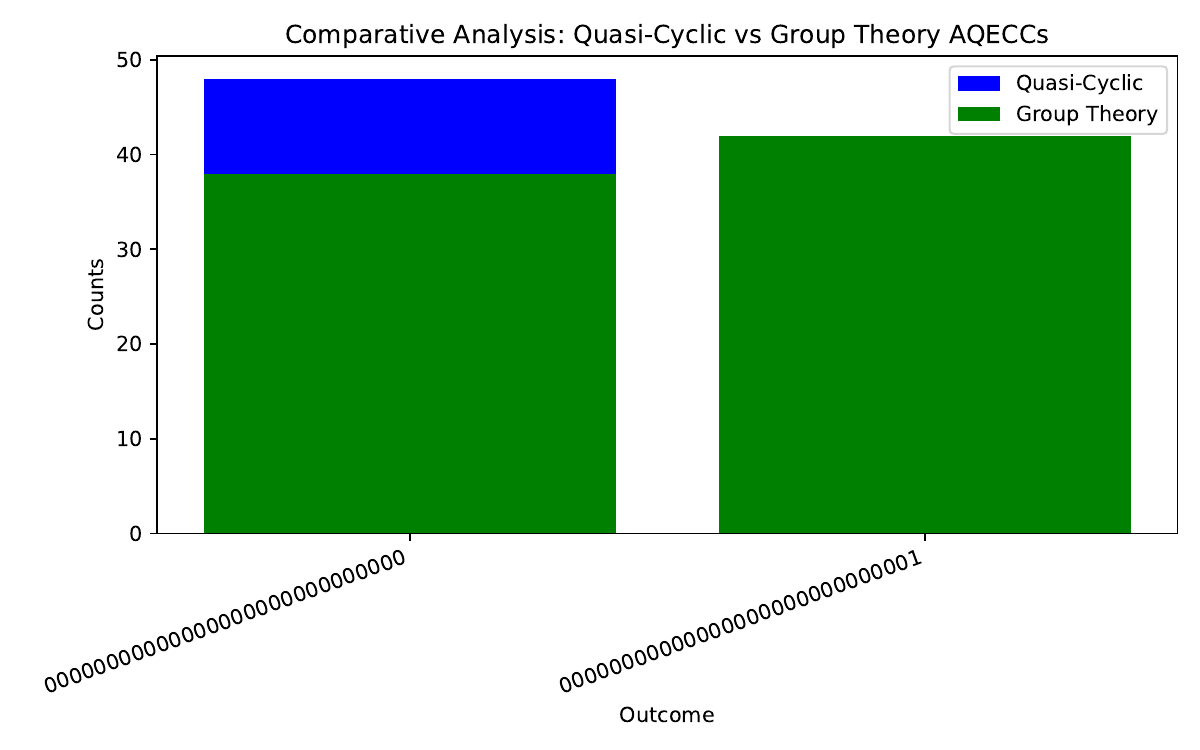}
    \caption{
    Comparative analysis between GT and QC
properties when transforming 1-qubit to 29-qubits under AQECCs.}
    \label{fig:20}
\end{figure}

Table~\ref{Table:8} shows the two numerical simulation results of AQECCs, using both QC and GT approaches, which contain high performance to perform and correct errors in the process of mapping 1-qubit to 29-qubit. 
In the tables ~\ref{Table:7} and \ref{Table:8}, we show only two outcomes with their respective counts, since the number of errors corrected exceeds the number of initial qubits. To generate a more complete set of results, an increase for $N$-qubits is necessary. Suppose the error correction facility is limited to correcting only 2 and 5 errors. In this case, the counts for outcomes with higher error rates may decrease significantly, following the properties of QC and GT. In QEC, increasing the number of physical qubits ($M$ in this case) and the ability to correct errors usually comes at the cost of increased complexity and resource requirements. 

Figure~\ref{fig:18} shows the analysis of initial and final quantum states by applying the $H$-gate while identifying 5 errors with the $X$-gate in the mapping process from 1-qubit to 29-qubits. Specifically, in Figure~\ref{fig:18}(a), the application of the $H$-gate to the quantum circuit generates a superposition of the qubit while simultaneously integrating $X$-gates to identify and correct the five errors in the final 29-qubit quantum circuit, as shown in Figure~\ref{fig:18}(b).

 Figure~\ref{fig:19} provides a visualization of the behavior of GT and QC under AQECCs during the transformation process from 1-qubit to 29-qubits, demonstrating the ability to correct five errors. The suitability of the proposed plot is evident from the graphical simulations shown in Figure~\ref{fig:19}(a) to Figure~\ref{fig:19}(b) for QC and GT properties, respectively.
 
Figure~\ref{fig:20} shows another comparison between the QC and GT properties of transforming 1-qubit to 29-qubits. In particular, the QC properties show suitability and effective performance in correcting five errors, as indicated by the tallest bar in this comparison plot compared to the bars representing the GT properties.  
 \end{enumerate}

Analysis of the count distribution provides valuable insight into the behavior of QC and GT under AQECCs in Tables~\ref{Table:5},\ref{Table:6},\ref{Table:7} and \ref{Table:8}. The different frequencies of the outcomes suggest a non-uniform distribution, indicating a higher probability of certain quantum states being realized. This non-uniformity is influenced by specific properties of the ECCs and the quantum interference introduced by the $H$-gates. This analysis improves our understanding of the quantum behavior of QC and GT in the context of AQECCs and contributes to the evaluation of their effectiveness in error correction and quantum stability.

The visual representation of the results for QC and GT (AQECCs) is shown in Figures~\ref{fig:10},\ref{fig:13},\ref{fig:16} and \ref{fig:19}, where each bar corresponds to a possible measurement result. The blue and green colors indicate the QC and GT, respectively, under the properties of AQECCs, and the count distribution reflects the associated probabilities for different measurement outcomes. In the mathematical formalism, if \( |\psi\rangle = \sum_{i=1}^{n} c_i|a_i\rangle \) represents the superposition state, then the probability of observing \( |a_i\rangle \) is determined by \( |c_i|^2 \). The effectiveness of the QC (AQECCs) in error correction is evaluated by the measured probabilities of different outcomes. This mathematical approach provides a comprehensive understanding of how quantum states and operations contribute to error correction in quantum computing, integrating matrices, state vectors, and principles from quantum mechanics and group theory.
The comparative analysis between QC and GT under AQECCs involves evaluating their performance in correcting errors and exploring common mathematical foundations. The quantitative evaluation considers counting distributions \(C_{\text{QC}}\) and \(C_{\text{GT}}\) for QC and GT properties, respectively, using linear bar graphs, where \(C_{\text{QC}}\) is the count of QC and \(C_{\text{GT}}\) is the count of GT. Deviations in these distributions may indicate variations in performance and susceptibility to quantum noise. The observed interference patterns highlight the quantum nature of both codes as influenced by $H$ gates. 
The common mathematical foundation using $H$ gates and measurement operations, consistent with the fundamental principles of QEC, is visually represented by a taller bar graph for a comprehensive understanding of their distinct characteristics and strengths.

\subsection{Statistical analysis}

Within the framework of the provided code and the simulated QEC, statistical measures~\cite{24} are essential to quantitatively evaluate the success and reliability of the error correction process. In assessing the effectiveness of transforming $N$-qubits into $M$-qubits equipped with the ability to correct $P$ errors through QO, QC, and GT properties. The primary statistical measure used is the error rate, which is calculated as the ratio of the total number of counts with errors to the total number of counts. The error rate is expressed as a percentage.
\begin{equation}
 \text{Error Rate}(\eta \%) = \frac{\sum \text{Counts with Errors}}{\sum \text{Total Counts}} \times 100 
\end{equation}
which provides a quantifiable indication of the accuracy and reliability of the transformation. A lower error rate means a more robust transformation, indicating improved error correction capabilities. Furthermore, the mean $\mu$ and the variance $\sigma^2$ are, respectively,
given by
\begin{align}
&\mu = \frac{{\sum \text{{counts}}}}{{\text{{number of outcomes}}}},
\\
&\sigma^2 = \frac{\sum_{i=1}^{n} (x_i - \mu)^2}{n}, 
\end{align}
where \(x_i\) is each count and \(n\) is the number of results. The counts in the experimental results provide insight into the distribution of the results. A higher mean count indicates a more successful error correction process, while a lower variance indicates greater consistency in the correction results. In general, lower error rates indicate better error correction performance. Consequently, for the $C_4$ transformation with QO and GT properties, the error correction performance appears to be more effective compared to other transformations with a more successful error correction process, while the $C_1$ and $C_2$ cases have greater consistency in the correction results. In contrast, the $C_3$ case proves to be less effective in error correction compared to the other cases, as shown in Table~\ref{tab: 1z}.

\begin{table}[!ht]
    \centering
    \caption{Analysis of successful error correction, consistency, and performance of four cases of the transformation of qubits with QO and GT under QOCCCs.}
    \label{tab: 1z}
    \begin{tabular}{|p{.8cm}|p{.8cm}|p{.8cm}|p{.8cm}|p{.8cm}|}
        \hline
    $C_n$ & $P$ & $\mu$ & $\sigma^2$ & $\eta \%$ \\ \hline
     $C_1$ & 1 & 3.2 & 1.16 & 68.75 \\
     $C_2$ &1 & 3.2 & 1.16 & 68.75 \\
    $C_3$ & 2 & 3.6 & 3.84 & 61.11 \\
    $C_4$ & 5 & 5.1 & 3.29 & 29.41 \\ \hline
    \end{tabular}
    
\end{table}
\noindent where $C_n$,with $n=1,...,4$ represents the four cases introduced in section \ref{section 3} .

\begin{table}[!ht]
\centering
\caption{Evaluate the effectiveness, reliability, and performance of error correction during the four cases using QC and GT properties within AQECCs.}
    \label{tab: 1p}
   \begin{tabular}{|c|c|c|c|c|c|c|c|}
    \hline 
   \multicolumn{1}{|c|}{  \(C_n\)} &
   \multicolumn{1}{|c|}{\(P\)} &
      \multicolumn{2}{c|}{$\mu$} &
      \multicolumn{2}{c|}{$\sigma^2$} &
      \multicolumn{2}{c|}{$\eta \%$} \\ 
      \hline
    \textbf{}& \textbf{} & GT & QC & GT & QC & GT & QC\\
    \hline
    $C_1$ &1 & 10 & 10 & 3.25& 6 & 83.75&77.50\\
      $C_2$ &1 & 5& 5 & 3.5 & 3.25 &87.50&88.75 \\
    $C_3$ &2 & 40& 40 & 32& 16 & 40.00& 45.00\\
    $C_4$ &5 &40 &40 &32 &2 &40.00 &47.50\\ \hline
  \end{tabular}
\end{table}

Table~\ref{tab: 1p} evaluates the efficiency, consistency, and overall performance based on mean, variance, and error rate in error correction during qubit transformation using QC and GT properties within AQECCs. As a result, case $C_4$, under QC properties, and process $C_3$, with both QC and GT, show superior error correction performance. In addition, both cases $C_4$ and $C_1$ show higher consistency in the correction results for QC and GT, respectively. In contrast, the $C_2$ case proves to be less effective in error correction compared to the other methods.

\section{Conclusion}\label{section 5}

We have successfully investigated the transformation of a small number of qubits into a larger number of qubits via the properties of GT, QO features (2D-QOCCCs), and QC attributes (AQECCs). We calculated and plotted the results using a simulation model based on the error correction capability under 2D-QOCCCs with the same behavior as QO and GT. We found that mapping a 1-qubit state to a 29-qubit configuration, correcting 5 errors, showed high performance, indicating a successful error correction process with comparable characteristics in both GT and QO scenarios via 2D-QOCCCs. Beyond theoretical formulations, practical simulations were performed to compare QC codes and GT methods for mapping qubits under AQECCs. The provided code examples demonstrated error correction using QC structures and $H$ gates, revealing different quantum behaviors in QC and GT through AQECCs. The results show that the 1-qubit to 13-qubit transformation exhibits high performance and successfully corrects two errors via AQECCs.

 In addition, the analysis of probability amplitudes in conjunction with counts provided a comprehensive understanding of the behavior of quantum states, shedding light on both frequency and probability in the context of 2D-QOCCCs and AQECCs due to the application of the $H$-gate. This thorough analysis advances our understanding of QEC and provides a basis for refining and optimizing these codes in future quantum applications.

\section*{Acknowledgment}

This material is based upon work supported by the Air Force Office of Scientific Research under award number FA2386-22-1-4062
\subsection*{ Conflict and Interest}
All Authors declared no conflicts of interest

\bibliographystyle{plain}

\begin{thebibliography}{9}

\bibitem{PhilosophicalTransactions}
Steane, A. M. "Introduction to quantum error correction." Philosophical Transactions of the Royal Society of London. Series A: Mathematical, Physical and Engineering Sciences 356, no. 1743: 1739-1758, 1998. \ \href{https://doi.org/10.1098/rsta.1998.0246}{doi:10.1098/rsta.1998.0246}

\bibitem{PhysicalreviewA52}
Shor, Peter W. "Scheme for reducing decoherence in quantum computer memory." Physical review A 52, no. 4 : R2493, 1995.\ \href{https://doi.org/10.1103/PhysRevA.52.R2493}{doi:10.1103/PhysRevA.52.R2493}

\bibitem{PhysRevLett}
A. R. Calderbank, E. M. Rains, P. W. Shor, and N. J. A. Sloane. Quantum error correction and orthogonal geometry. Phys. Rev. Lett., 78:405–408, 1997.\ \href{https://doi.org/10.1103/PhysRevLett.78.405}{doi:10.1103/PhysRevLett.78.405.}

\bibitem{ProceedingsLondon}
Calderbank, A. Robert, Peter J. Cameron, William M. Kantor, and Jaap J. Seidel. "$\mathbf{Z}$4-Kerdock codes, orthogonal spreads, and extremal Euclidean line-sets." Proceedings of the London Mathematical Society 75, no. 2: 436-480, 1997.\ \href{ https://doi.org/10.1112/S0024611597000403 }{doi:10.1112/S0024611597000403 }
\bibitem{ResearchPapers}
Dávideková, Monika. "Generalized construction of two-dimensional quasi-complete complementary codes." Research Papers Faculty of Materials Science and Technology Slovak University of Technology 21.,12-17, Special-Issue, 2013. \ \href{https://doi.org/10.2478/rput-2013-0003}{doi:10.2478/rput-2013-0003}

\bibitem{IEEEWireless}
Sharma, Naresh, and Constantinos B. Papadias. "Improved quasi-orthogonal codes." In 2002 IEEE Wireless Communications and Networking Conference Record. WCNC 2002 (Cat. No. 02TH8609), vol. 1, pp. 169-171. IEEE, 2002.\ \href{ https://doi.org/10.1109/WCNC.2002.993484}{doi:10.1109/WCNC.2002.993484}

\bibitem{IEEETransactionscommunications}
Jafarkhani, Hamid. "A quasi-orthogonal space-time block code." IEEE Transactions on Communications 49.1: 1-4, 2001.\ \href{ https://doi.org/10.1109/26.898239}{doi:10.1109/26.898239}

\bibitem{IEEETransactionsInformation}
Su, Weifeng, and Xiang-Gen Xia. "Signal constellations for quasi-orthogonal space-time block codes with full diversity." IEEE Transactions on Information Theory 50, no. 10: 2331-2347, 2004.\ \href{https://doi.org/ 10.1109/TIT.2004.834740}{doi:10.1109/TIT.2004.834740}

\bibitem{ieeexplore.ieee.org}
Turcsany, Matúš, and Peter FarkaS. "Two-dimensional quasi-orthogonal complete complementary codes." In SympoTIC'03. Joint 1st Workshop on Mobile Future and Symposium on Trends in Communications., pp.37-40. IEEE, 2003. \ \href{https://doi.org/ 10.1109/TIC.2003.1249083}{doi:10.1109/TIC.2003.1249083}

\bibitem{7}
Lv, Jingjie, Ruihu Li, and Yu Yao. "Quasi-cyclic constructions of asymmetric quantum error-correcting codes." Cryptography and Communications 13.5:661-680, 2021. \ \href{ https://doi.org/10.1007/s12095-021-00489-9}{ doi:10.1007/s12095-021-00489-9}
\bibitem{arXivpreprint}
Aly, Salah A. "Constructing Asymmetric Quantum and Subsystem Cyclic Codes." arXiv preprint arXiv:0906.5339, 2009.\ \href{https://doi.org/10.48550/arXiv.0906.5339}{doi:10.48550/arXiv.0906.5339}
\bibitem{IEEEAccess}
Galindo, Carlos, Fernando Hernando, Ryutaroh Matsumoto, and Diego Ruano. "Asymmetric entanglement-assisted quantum error-correcting codes and BCH codes." IEEE Access 8: 18571-18579, 2020.\ \href{ https://doi.org/10.1109/ACCESS.2020.2967426}{doi:10.1109/ACCESS.2020.2967426}

\bibitem{SpringerBerlinHeidelberg}
Just, Bettina. "Quantum Gates on One Qubit." Quantum Computing Compact: Spooky Action at a Distance and Teleportation Easy to Understand. Berlin, Heidelberg: Springer Berlin Heidelberg, . 69-82, 2023.\ \href{https://doi.org/10.1007/978-3-662-65008-0-9}{doi:10.1007/978-3-662-65008-0-9}

\bibitem{25}
Bombin, Hector. "Clifford gates by code deformation." New Journal of Physics 13, no. 4: 043005, 2011.\ \href{http://doi.org/10.1088/1367-2630/13/4/043005}{doi:10.1088/1367-2630/13/4/043005}

\bibitem{26}
Moses, Steven A., Charles H. Baldwin, Michael S. Allman, R. Ancona, L. Ascarrunz, C. Barnes, J. Bartolotta et al. "A race-track trapped-ion quantum processor." Physical Review X 13, no. 4: 041052, 2023.\ \href{https://doi.org/10.1103/PhysRevX.13.041052}{doi:10.1103/PhysRevX.13.041052}

\bibitem{cyberleninka.ru}
Senashov, V. I. "Properties of locally cyclic groups."Siberian Journal of Science and Technology.Vol. 18, No. 2, P.290–293, 2017.\ \href{ https://journals.eco-vector.com/2712-8970/article/view/503306}{doi:journals.eco-vector.com/2712-8970/article/view/503306}

\bibitem{27}
Planat, Michel. "On the cyclotomic quantum algebra of time perception." arXiv preprint quant-ph/0403020, 2004. \ \href{https://doi.org/10.48550/arXiv.quant-ph/0403020}{doi:10.48550/arXiv.quant-ph/0403020}
\bibitem{28}
Pourkia, Arash, and J. Batle. "Cyclic groups and quantum logic gates." Annals of Physics 373: 10-27, 2016. \ \href{https://doi.org/10.1016/j.aop.2016.06.023}{doi:10.1016/j.aop.2016.06.023}
\bibitem{AnnalidiMatematicaPuraedApplicata}
Dvornicich, E., and Marco Forti. "Paracyclic and quasi-cyclic groups." Annali di Matematica Pura ed Applicata 145, no. 1 : 297-316, 1986.\ \href{https://doi.org/10.1007/BF01790544}{doi:10.1007/BF01790544}


\bibitem{32}
Grier, Daniel, and Luke Schaeffer. "The Classification of Clifford Gates over Qubits." Quantum 6: 734, 2022.\ \href{https://doi.org/10.22331/q-2022-06-13-734}{doi:10.22331/q-2022-06-13-734}
\bibitem{33}
Dilley, Daniel, Alvin Gonzales, and Mark Byrd. "Identifying quantum correlations using explicit SO (3) to SU (2) maps." Quantum Information Processing 21, no. 10: 343, 2022.\ \href{ https://doi.org/10.1007/s11128-022-03679-3}{doi:10.1007/s11128-022-03679-3}





\bibitem{Entropy25}
Li, Yuan, and Jin-Yang Li. "Quantum Coding via Quasi-Cyclic Block Matrix." Entropy 25, no. 3: 537, 2023. \ \href{https://doi.org/10.3390/e25030537}{ doi:10.3390/e25030537}

\bibitem{IEEETransactionsonInformationTheory13}
Townsend, Richard, and E. Weldon. "Self-orthogonal quasi-cyclic codes." IEEE Transactions on Information Theory 13, no. 2: 183-195, 1967.\ \href{https://doi.org/10.1109/TIT.1967.1053974}{doi:10.1109/TIT.1967.1053974}

\bibitem{34}
Nguyen, Van-Linh, Po-Ching Lin, Bo-Chao Cheng, Ren-Hung Hwang, and Ying-Dar Lin. "Security and privacy for 6G: A survey on prospective technologies and challenges." IEEE Communications Surveys \& Tutorials 23, no. 4: 2384-2428, 2021.\ \href{https://doi.org/ 10.1109/COMST.2021.3108618}{doi: 10.1109/COMST.2021.3108618}

\bibitem{19}
Long, Junling, Tongyu Zhao, Mustafa Bal, Ruichen Zhao, George S. Barron, Hsiang-sheng Ku, Joel A. Howard et al. "A universal quantum gate set for transmon qubits with strong ZZ interactions." arXiv preprint arXiv:2103.12305, 2021.\ \href{https://doi.org/10.48550/arXiv.2103.12305}{doi:10.48550/arXiv.2103.12305}

\bibitem{29}
Shcherbacov, Victor. Elements of quasigroup theory and applications. Chapman and Hall/CRC, 2017.\ \href{ https://doi.org/10.1201/9781315120058 }{doi:10.1201/9781315120058 }

\bibitem{20}
Buehrer, R. Michael. Code division multiple access (CDMA). Springer Nature, 2022.\ \href{https://doi.org/10.1007/978-3-031-01673-8}{doi:10.1007/978-3-031-01673-8}

\bibitem{21}
Zigangirov, Kamil Sh. Theory of code division multiple access communication. John Wiley \& Sons, 2004.\ \href{https://doi.org/10.1002/047165549X}{doi:10.1002/047165549X}

\bibitem{22}
Brooks, Michael, ed. Quantum computing and communications. Springer Science \& Business Media, 2012.\
\href{https://doi.org/10.1007/978-1-4471-0839-9}{doi:10.1007/978-1-4471-0839-9}

\bibitem{23}
Ioffe, Lev, and Marc Mézard. "Asymmetric quantum error-correcting codes." Physical Review A 75, no. 3: 032345, 2007. \ \href{https://doi.org/10.1103/PhysRevA.75.032345}{doi:10.1103/PhysRevA.75.032345}

%




\bibitem{35}
 Bartz, H. \& Wachter-Zeh, A. Improved decoding and error floor analysis of staircase codes. Des. Codes Cryptogr. 87, 647–664, 2019. \ \href{https://doi.org/10.1007/s10623-018-0587-x}{doi:10.1007/s10623-018-0587-x}
 \bibitem{36}
 Ge, Yan, Wu Wenjie, Chen Yuheng, Pan Kaisen, Lu Xudong, Zhou Zixiang, Wang Yuhan, Wang Ruocheng, and Yan Junchi. "Quantum Circuit Synthesis and Compilation Optimization: Overview and Prospects." arXiv preprint arXiv:2407.00736 , 2024.\ \href{https://doi.org/10.48550/arXiv.2407.00736
}{doi:10.48550/arXiv.2407.00736}











  
\end{thebibliography}

 \end{document}